\documentclass[fleqn,usenatbib]{mnras}

\usepackage{newtxtext,newtxmath}
\usepackage[T1]{fontenc}

\DeclareRobustCommand{\VAN}[3]{#2}
\let\VANthebibliography\thebibliography
\def\thebibliography{\DeclareRobustCommand{\VAN}[3]{##3}\VANthebibliography}

\usepackage{graphicx}	% Including figure files
\usepackage{amsmath}	% Advanced maths commands
\usepackage{hyperref}
\usepackage{xurl}
\hypersetup{breaklinks=true}
\usepackage{breakurl}

\title[NGC 315: UV star formation]{Enhanced UV emission knot in the giant radio galaxy NGC 315: Hint of patchy star formation?}

\author[B. Ananthamoorthy et al.]{
Bannanje Ananthamoorthy,$^{1}$
Debbijoy Bhattacharya,$^{1}$\thanks{E-mail: debbijoy.b@manipal.edu}
Dipanjan Mukherjee$^{2}$
and P. Sreekumar$^{1}$
\\
$^{1}$Manipal Centre for Natural Sciences, Manipal Academy of Higher Education, Karnataka, Manipal, 576 104, India\\
$^{2}$Inter-University Centre for Astronomy and Astrophysics, Pune- 411007, India\\
}

\date{Accepted XXX. Received YYY; in original form ZZZ}

\pubyear{\the\year{}}

\begin{document}
\label{firstpage}
\pagerange{\pageref{firstpage}--\pageref{lastpage}}
\maketitle

\begin{abstract}
High-resolution {\sl AstroSat}-UltraViolet Imaging Telescope (UVIT) observations revealed a knot of UV emission, $\sim 1.7$ kpc away from the centre of NGC 315, a nearby elliptical galaxy hosting a giant (Mpc scale) radio source with a jet. We suggest that this patchy and spatially extended UV emission is likely due to ongoing star formation (SF) in the galaxy. The estimated SF rate (SFR) averaged over 100 Myr for the UV knot ($0.23\pm0.10$ M$_{\odot}$ yr$^{-1}$) is significantly higher compared to a typical elliptical galaxy. As the galaxy does not show the signatures of recent major mergers, the possible mechanisms for the triggered SF include AGN feedback or minor mergers. {\sl Hubble Space Telescope} ({\sl HST}) observations reveal dust filaments that extend through a UV knot. The origin of dusty filaments, though not clear, could be associated with gas clouds as a result of a minor merger, cooled gas falling into the central BCG and/or condensing of the gas uplifted by AGN jet. No significant clumpy UV emission is observed in other regions along the dust filament. We speculate that mechanical feedback from the AGN jet could be playing a role in triggering SF in the UV knot.
\end{abstract}


%%%%%%%%%%%%%%%%%%%%%%%%%%%%%%%%%%%%%%%%%%%%%%%%%%

%%%%%%%%%%%%%%%%% BODY OF PAPER %%%%%%%%%%%%%%%%%%

\section{Introduction} \label{sec:intro}

Elliptical galaxies, characterised by their old stellar populations, are generally expected to exhibit faint Ultraviolet (UV) emission. However, UV emission is widespread in elliptical galaxies \citep[e.g.,][and references therein]{Connel1999}. Observations from the {\sl HST} have directly detected young stars in several nearby elliptical galaxies \citep{Ford2013}. Although the estimated star formation rate (SFR) is relatively low, on the order of $\sim 10^{-4} M_{\odot}$yr$^{-1}$, the presence of young stars and star clusters suggests that many elliptical galaxies are not entirely ‘red and dead' \citep{Ford2013, Tamhane2025}. While clumpy UV emission is often associated with star formation (SF), the underlying mechanism that triggers this SF remains unclear. 

Mergers could be one of the primary drivers of SF in elliptical galaxies. Simulations have suggested that both major and minor mergers can trigger SF in the galaxy \citep[e.g.,][]{Matteo2007, Hani2020}. Evidence for possible merger-driven SF exists in both low- and intermediate-redshifts \citep[e.g.,][]{Yi2005, Kaviraj2011}. A strong correlation is observed between UV emissions and the presence of morphological disturbances in elliptical galaxies \citep[e.g.,][]{Kaviraj2011}, suggesting a merger-induced SF. It is also noticed that in merger regions, the induced SF can be efficient with a lower gas depletion timescale $\left (\tau_{\text{dep}} = \frac{\text{gas surface density}~(\Sigma_{\text gas})}{\text{SFR}} \right )$ of $\sim 0.1-0.5$ Gyr in contrast to $\sim 2$ Gyr for typical SF regions \citep[e.g.,][]{Gracia2020}. 

Feedback from `Active Galactic Nuclei (AGN)' could be another prime driver of SF in elliptical galaxies. The AGN feedback is included in most of the galaxy simulation models \citep[see, e.g.,][for review]{Oppenheimer2021}. Negative feedback from strong radio sources is often invoked to explain the lack of significant recent SF in massive ellipticals and to explain the cutoff at higher mass in the galaxy mass function \citep[e.g.,][]{Croton2006, Shabala2009}. Though AGN jet has the energetics to influence the properties of the galaxy, the mechanism through which they can affect the galaxy's SF is still not clear \citep[][and references therein]{Hardcastle2020, Fabian2012, Harrison2024}. Simulations suggest that the increased pressure of embedded gas clouds due to jet mode AGN feedback can facilitate SF \citep{Fragile2004, Fragile2017, Wagner2012, Silk2013}. Recent simulations have suggested that though SF can occasionally increase, in the long run, and on a global scale, SF will be suppressed \citep[e.g.,][]{Mondal2021, Mukherjee2018, Mukherjee2018a, Delogu2008}. \citet{Mukherjee2018} also suggested that the nature of feedback, either positive (enhanced SF) or negative (suppression of SF), depends on the jet power, orientation, and density of the gas.

One of the prominent observational evidence of AGN jets influencing the environment is through X-ray observations. The cavities are observed in the X-ray emitting hot cluster medium along the direction of the radio jet \cite[e.g.,][]{Boehringer1993, Dunn2004}. This suggests that AGN jets expel gas as they propagate in the cluster environment \cite[e.g.,][]{Fabian2012}. However, observational evidence for its direct impact on SF is limited to a few galaxies. In Centaurus A (Cen. A), star-forming filaments are observed along the outer edge of the jet, which is likely to be triggered by the AGN jet \citep{Crockett2012, Salome2016, Santoro2015, Joseph2022}. Observations have also suggested for possible AGN jet-triggered SF in Minkowski object along the jet of NGC 541 \citep{Croft2006, Lacy2017}, radio galaxy B2 0258+35 \citep{Murthy2019}, etc. Alignment between radio and optical structures is also observed in high-redshift radio galaxies, most likely due to positive AGN feedback \citep[e.g.,][]{Chambers1987, McCarthy1993, Pentericci2001, Zirm2005, Lacy1999}. \citet{Duggal2024} observed UV enhancements in Compact Steep Spectrum (CSS) radio sources at low to intermediate redshifts using high-resolution NUV observations from {\sl HST}. These UV enhancements have also been attributed to jet-triggered SF in those galaxies.

Far-UV (FUV) emissions are observed in many brightest cluster galaxies (BCGs) having cool cores \citep{Hu1992}. {\sl HST} observations have indicated the presence of many bright UV knots with significant ongoing SF (even extending up to 100 M$_\odot$ yr$^{-1}$) at the centres of the BCGs \citep[e.g.,][]{Koekmoer1999, Odea2004, Odea2010, Russel2017, Fogarty2015, Donahue2015, Arsen2024}. \citet{Odea2004} further observed that SF is enhanced along the edges of the radio sources in A1795 and A2597, suggesting that shocks from the radio source may help trigger SF. \citet{Donahue2015} noticed that the UV excess BCGs at $0.2 < z < 0.9$ exhibit distinct UV morphologies with knots and filaments, in contrast to the smooth UV profiles of quiescent BCGs. They propose that such features correspond to the streams of clumpy SF in these galaxies. \citet{Kolokythas2022} reported that, in nearby galaxy groups, only a small fraction ($\sim 13\%$) of central ellipticals show substantial SF.

The hot gas environments of BCGs have also been extensively studied in X-rays \citep[e.g.,][]{Chen2007, Sun2009, Osullivan2017, Babyk2018}. \citet{Sun2009}, suggested that essentially all the BCGs with strong radio AGN ($1.4$ GHz luminosity $> 2 \times 10^{23}$ W Hz$^{-1}$) have X-ray cool cores (with central cooling time of $<1$ Gyr). Using X-ray and IR observations, \citet{Odea2008} argued that the heating mechanisms in cool cores are efficient in suppressing the gas cooling but do not halt it entirely. The ``precipitation-driven AGN feedback'', where AGN jets uplift low-entropy gas that subsequently condenses and forms stars, has also been proposed to explain the observed UV star-forming structures in BCGs \citep[e.g.,][]{Donahue2015}.

{\sl HST} observations have revealed the presence of dust lanes in most of the elliptical galaxies within hundreds of parsecs to a kpc of the nucleus \citep{Dokkum1995}.  Although dust lanes are common in early-type galaxies, the origin of the gas/dust content in the elliptical galaxy is still debatable. \citet{Dokkum1995} estimated that the mass of dust ranged from $10^{3}$ to $10^{7} M_{\odot}$ and suggested that the dust lanes in these galaxies could be due to an external origin, such as mergers.

\citet{Finkelman2012} used optical H$\alpha$ and near-IR observations and suggested that ionizing gas and dust have a common origin and are well mixed. According to them, the dust morphology observed in elliptical galaxies is difficult to explain purely based on internal origins, such as mass loss due to older stellar populations, indicating that external factors like mergers are responsible for the dust and gas. However, they also suggested that outflow disruption from the central AGN could be responsible for irregular dust features observed in the galaxy. The possible role of AGN in dust formation has also been suggested by several other studies \citep[e.g.,][]{Temi2007, Richtler2020}.

\vspace{0.2cm}
\noindent{\bf NGC 315: Galaxy overview}
\vspace{0.1cm}

NGC 315 is a nearby elliptical galaxy \citep[redshift, z = 0.01648;][]{Trager2000} having a size of $1.6^{\prime}$ \citep{RC3} along semi-major axis (in optical band) and located at RA = 00h57m48.883s, DEC = 30d21m08.812s \citep[NED;][]{Fey2004}\footnote{\url{https://ned.ipac.caltech.edu/}}. It is at the centre of a giant radio galaxy of Fanaroff-Riley type I \citep[FR1;][]{Fanaroff1974} extending up to $1.7$ Mpc \citep[]{Bridle1976}. The central AGN is detected across the entire electromagnetic spectrum, from radio to $\gamma$-rays \citep[e.g.,][]{Tomar2021}. 

The radio jet has been extensively studied over small (milli arcsec) to large (Mpc) scales \citep[e.g.,][]{Bridle1976, Bicknell1986, Venturi1993, Bicknell1994, Cotton1999, Canvin2005, Liang2006, Boccardi2021, Park2021, Ricci2022, Ricci2025}. The radio jet is aligned along the minor axis of the optical galaxy \citep{Bridle1976}. It is suggested that the jet in NGC 315 is collimated by the pressure of the surrounding medium \citep{Park2021}. \citet{Bicknell1994} estimated the minimum energy flux of the jet to be $\sim 10^{43}$ergs s$^{-1}$ using the conservation laws for the entraining jet. Based on the estimation of jet power using various methods, \citet{Ricci2022} adopted a value of $1.4\times10^{44}$ ergs s$^{-1}$ for NGC 315, which is an order of magnitude higher than the minimum energy flux derived by \citet{Bicknell1994}. The velocity profile, magnetic field, and spectral index maps have also been studied in great detail for this galaxy from pc to kpc scales \citep{Liang2006, Ricci2025}. An inner jet extending up to the 10 kpc scale is also observed in the X-ray band of {\sl Chandra} observation \citep[e.g.,][]{Worrall2003, Worrall2007}.

NGC 315 is the BCG of the NGC 315 group, which consists of 23 member galaxies \citep[NGC 315 group;][]{Chen2012, Miller2002}. Based on X-ray observations, NGC 315 is classified as a cool core galaxy \citep{Sun2009}. \citet{Chen2012} noticed that the X-ray luminosity of the NGC 315 group is lower than that expected from its galaxy velocity dispersion, implying a relatively low extended hot gas density in this group. \citet{Morganti2009} detected five galaxies with HI emission in the NGC 315 group, upon which they inferred that the NGC 315 group is a gas-rich system.

Mid-infrared colour analysis by \citet{Kolokythas2022} suggests possible dust-obscured SF in NGC 315. They notice that NGC 315 is the only system in their sample identified as star-forming that resides in a group with a hot inter-group medium \citep{Osullivan2017}. Using {\sl Galaxy Evolution Explorer (GALEX)} FUV observations, \citet{Kolokythas2022} estimated the FUV SFR, averaged over 100 Myr, to be 0.255 M$_\odot$ yr$^{-1}$. However, this estimate does not include corrections for internal dust attenuation or the contribution from the old stellar population.

{\sl HST} observations have revealed a dust disc of 820 pc in diameter in the galaxy's central region \citep{Kleijn1999}. Several patches of dust are also observed along the southwest of the nucleus up to 5$^{\prime\prime}$ \citep[1.5 kpc; ][]{Kleijn1999}. \citet{Morganti2009} observed a broad and narrow spectral component of HI absorption in the central region. They attribute the broad component to the infalling gas into the central AGN. However, they do not entirely rule out the possibility of gas entrainment by the radio jet. The narrow components are not close to the AGN, and they associate it with clouds similar to dusty patches observed in HST observations.

 As the galaxy is nearby and hosts large-scale radio jets, it is a very good candidate to study the AGN jet impact on the host galaxy's SF properties. In this work, we explore the UV emission in the central region of NGC 315 using UVIT onboard {\sl AstroSat}. We further explore dust distribution in the central region of the galaxy and its possible connection with SF and AGN activity. We consider `$\Lambda$CDM cosmology' with $\Omega_m$ = 0.308, $\Omega_{vac}$ = 0.692, and H$_{0}$ = 67.8 km s$^{-1}$ Mpc$^{-1}$ \citep{Plank2016}. The redshift of 0.01648 corresponds to a `luminosity distance' of $72.3$ Mpc\footnote{\url{https://www.astro.ucla.edu/~wright/CosmoCalc.html}}, an `angular size distance' of $\sim 70$ Mpc, and an angular scale of $\sim 340$ pc arcsec$^{-1}$.

%%%%%%%%%%%%%%%%%%%%%%%%%DATA%%%%%%%%%%%%%%%%%%%%%%%%%%%%%%%%%%%%%

\section{Data}
\subsection{{\sl AstroSat}-UVIT}
UVIT onboard {\sl AstroSat} comprised of two `Ritchey-Chretien' telescopes. These telescopes are designed to carry out observations in FUV (130-180 nm) and Near-UV (NUV) (200-300 nm) with a `field of view (FOV)' of $14^{\prime}$ radius. Each channel has a $512 \times$512 CMOS detector, which reads $\sim$29 frames per second \citep{Tandon2017}. The event positions are computed with a precision of 1/32 of a pixel on the $512 \times 512$ CMOS detector \citep{Tandon2017}.

Level-1 data of our observations of NGC 315 (Observation ID: A11\_101T01\_9000005238; Observation date: July 14, 2022) are obtained from the data repository of the `Indian Space Science Data Center (ISSDC)'\footnote{\url{https://astrobrowse.issdc.gov.in/astro_archive/archive/Home.jsp}} and utilized the \texttt{CCDLAB} \citep{Postma2017, Postma2021} pipeline to obtain the Level-2 images and exposure maps. We have observations in the FUV BaF2 filter (F154W) with UVIT. \texttt{CCDLAB} is used to correct for spacecraft drift, field distortions, flat field correction, and cosmic-ray corrections to provide the counts image and corresponding exposure map \citep[see][for details]{Postma2017, Postma2021}. The final image has a plate scale of $\sim 0.416^{\prime \prime}$ per pixel, angular resolution of $\sim1.3^{\prime \prime}$, and exposure time of 25.7ks. 

\subsection{{\sl Swift}-UVOT}
NUV observations are obtained from {\sl Swift} UVOT instrument \citep{Roming2005} in {\it uvw1}, {\it uvw2}, and {\it uvm2} filters. The raw data (observations spanned from year 2007 to 2017) are obtained from NASA’s High Energy Astrophysics Science Archive Research Center (HEASARC)\footnote{\url{https://heasarc.gsfc.nasa.gov/cgi-bin/W3Browse/swift.pl}}. 
We utilised `\texttt{HEASOFT}' version 6.31.1 and UVOT `\texttt{CALDB}' version 20240201 for analysis. As NGC 315 is an extended source, `Data Reduction of Extended {\sl Swift} Sources Code (\texttt{DRESSCode})'\footnote{\url{https://spacetelescope.github.io/DRESSCode/}} tool is used for the analysis \citep{Decleir2019a, Decleir2019b}. \texttt{DRESSCode} is a fully automated pipeline that reduces {\sl Swift} UVOT images of extended sources. It is used for aspect correction, creating auxiliary maps, and combining separate frames. It also applies various corrections such as coincidence loss correction, large-scale sensitivity correction, and zero-point correction \citep{Decleir2019a, Decleir2019b}. The flux-calibrated final images have an angular resolution of $\sim3^{\prime \prime}$ in {\it uvw1}, {\it uvw2}, and {\it uvm2}, with average exposure times of $\sim 2.7$, $\sim 5$, and $\sim 4.1$ ks, respectively, in the central $10 ^{\prime \prime}$ region of NGC 315.

\subsection{{\sl GALEX}} 
Level-2 science-ready archival photometry observations from {\sl GALEX} \citep{Martin2005, Morrissey2007} in the FUV ($\lambda_{eff} = 1538.6\text{\AA}$) and NUV ($\lambda_{eff} = 2315.7\text{\AA}$) bands  are obtained from the MAST portal\footnote{\url{https://galex.stsci.edu/gr6/}}. The calibrated intensity and effective exposure map images with an angular resolution of 4.5-5.5$^{\prime\prime}$ are used for the analysis \citep{Morrissey2007}. Two observation tiles (GI2\_019001\_NGC315 and NGA\_NGC0315) with an exposure time of each $\sim 1.6$ ks were present for NGC 315. These observations were combined using the \texttt{reproject} routine \citep{Robitaille2018} in \texttt{Python}.

\subsection{{\sl HST}} %and {\sl Spitzer}-IRAC }}

We utilised the archival calibrated, drizzled {\sl HST} images ({\it Hubble} Advanced Products) obtained from the MAST portal\footnote{\url{https://mast.stsci.edu/portal/Mashup/Clients/Mast/Portal.html}}. We utilized F555W and F814W 
filter observations from the WFPC2 instrument \citep{Trauger1994, Holtzman1995} to understand the dust distribution in the central region of NGC 315.  The exposure time of the combined images were 460s in each filter and has a plate scale of $0.046^{\prime \prime}$ per pixel. The flux from Data Number (DN) is derived using the `PHOTFLAM' and `PHOTPLAM' parameters in the image header\footnote{\url{https://www.stsci.edu/hst/wfpc2/Wfpc2_dhb/WFPC2_longdhb.pdf}}.

\begin{figure*}
\centering
	\includegraphics[width=18cm]{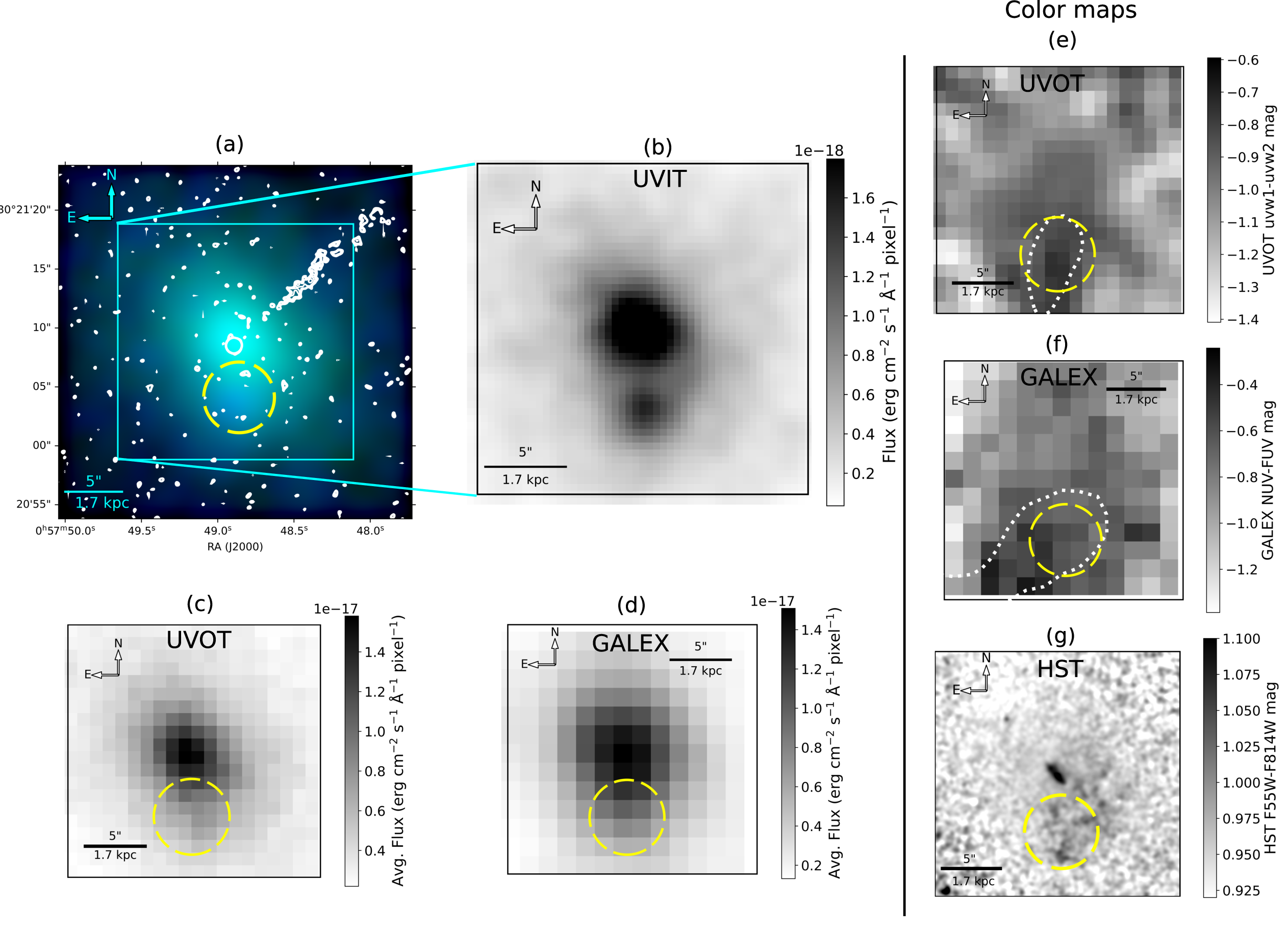}
    \caption{Panel (a): RGB composite colour image of the central 25$^{\prime\prime}$ region of NGC 315 with FUV (from {\sl AstroSat}-UVIT) in blue, NUV (from uvw1 filter of {\sl Swift}-UVOT) in green, and optical (from HST F555W filter) in red. Solid white contours correspond to the radio emission observed with 1.4 GHz VLA observations (obtained from the VLA archive), representing the direction of the jet. The dashed yellow circle corresponds to a $3^{\prime \prime}$ region centred at the knot of UV emission in all panels. The cyan box corresponds to a box of width 20$^{\prime\prime}$ centred at NGC 315. Panels (b)–(g) show flux or colour maps of the central $10^{\prime\prime}$ region. Panel (b): FUV F154W filter. Panel (c): {\sl Swift}-UVOT stacked image in $uvw1$, $uvw2$, and $uvm2$ filters. Panel (d): {\sl GALEX} stacked image in FUV and NUV filters. Panel (e): {\sl Swift} UVOT {\it uvw1}-{\it uvw2} color map. The dotted white contour corresponds to a colour of -0.87, which is a $1\sigma$ contour level obtained by fitting a 2D Gaussian around the peak value. Panel (f): {\sl GALEX} NUV-FUV colour map. The dotted white contour corresponds to a colour of -0.73, which is a $1\sigma$ contour level obtained by fitting a 2D Gaussian around the peak value. Panel (g): {\sl HST} F555W-F814W colour map.}
    \label{fig:UVIT_knot}
\end{figure*}

%%%%%%%%%%%%%%%%%%%%%%%%%%%%%%%%%%%Analysis and results%%%%%%%%%%%%
\section{Analysis and Results}

The RGB colour image of the central 25$^{\prime\prime}$ region of NGC 315 with FUV (from {\sl AstroSat}-UVIT) in blue, NUV (from uvw1 filter of {\sl Swift}-UVOT) in green, and optical (from {\sl HST} F555W filter) in red is provided in panel (a) of Fig.~\ref{fig:UVIT_knot}. The contours from 1.4 GHz observations from ‘Very Large Array (VLA)’\footnote{obtained from \url{https://www.vla.nrao.edu/astro/nvas/}} are overplotted to indicate the jet axis. AstroSat UVIT observations have $\sim 3$ times better resolution than {\sl GALEX} \citep{Morrissey2007}. The higher-resolution UVIT image reveals a knot of UV emission (Position: RA = 00h57m48.8730s, DEC = 30d21m04.377s) towards the southern direction $\sim 5^{\prime\prime}$ ($\sim 1.7$ kpc) away from the UV core, as shown in panel (b) of Fig.~\ref{fig:UVIT_knot}. The position of the UV knot is notably off the jet axis.

No NUV observations were available with {\sl AstroSat}-UVIT as the NUV channel ceased its operations in August 2018, and our observations were carried out in July 2022. Consequently, we have utilised {\sl Swift}-UVOT and {\sl GALEX} observations to search for NUV counterparts of the FUV knot. The stacked observations from UVOT ({\it uvw1} and {\it uvw2} filters) and {\sl GALEX} (FUV and NUV filters) are provided in panels (c) and (d) of Fig.~\ref{fig:UVIT_knot}, respectively. Unlike UVIT, the knot is not resolved in UVOT and {\sl GALEX} observations due to lower angular resolution. The excess FUV emission is pronounced in the colour map derived using {\sl Swift}-UVOT ({\it uvw1}- {\it uvw2}) observations as presented in panel (e) of Fig.~\ref{fig:UVIT_knot}. {\sl GALEX} colours (NUV-FUV) also show excess FUV towards the position of the UV knot. However, due to lower angular resolution, the position is not constrained well in the {\sl GALEX} colour map. The estimated colour in an aperture of radius $1.5^{\prime\prime}$ for the central and the knot region from UVOT observations was $-0.92\pm0.05$ and $-0.8\pm0.08$, respectively. Similarly, the colour obtained from {\sl GALEX} observations in an aperture of radius $3^{\prime\prime}$ was $-0.75\pm0.06$ and $-0.64\pm0.07$ for the central and knot region, respectively. These colours support the presence of excess FUV towards the position of the UVIT knot.

The source was observed by UVOT in 2007 and 2017. Therefore, we also generated the light curve by extracting the flux in an aperture with a radius of $1.5^{\prime\prime}$ centred on the UV knot position in {\it uvw1}, {\it uvw2}, and {\it uvm2} filters. We do not observe any significant variability in the source UV emission. The lightcurves and statistical test results are provided in Fig.~\ref{fig:lc_uvot} and Table~\ref{tab:var_test}, respectively, in Appendix~\ref{sec:var_test}.

No optical counterpart for the knot was detected in high-resolution {\sl HST} F555W and F814W filter observations. However, we have utilised HST observations to characterise the dust properties around the UVIT knot. The HST F555W-F814W colour map is provided in panel (g) of Fig. 1, which reveals the central dust disc along with several dust patches around the position of the UV knot.

\subsection{Determination of FUV flux from the UVIT knot}\label{UV_knot}
To obtain the flux from the UV knot, we fitted the galaxy emission in UVIT F154W band using \texttt{GALFIT} \citep{Peng2002, Peng2010}. We used the central $\sim 40^{\prime\prime} \times 40^{\prime \prime}$ ($100 \times 100$ pixels) region, which corresponds to the $\sim 0.2$ optical radius (R$_{25}$), to fit the galaxy emission. The sky value and the sky standard deviation (std) are obtained using the sliding average method. We considered a sliding box of size $32\times32$ pixels over the region of $600\times600$ pixels centred around the galaxy. The mean background (0.522 counts/pixel) and the mean std (0.935 counts/pixel) in these sliding boxes are considered for analysis. We noticed that if we add error due to source count (Poisson error) at each pixel in quadrature with sky std to calculate the sigma map (as in Section~8 of the \texttt{GALFIT} manual\footnote{\url{https://users.obs.carnegiescience.edu/peng/work/galfit/README.pdf}}), the derived \texttt{GALFIT} model significantly and systematically underestimates the observed flux. This could be due to the low counts/pixels in the UVIT observations. To understand this effect, we simulated the galaxies having counts similar to those in the current observation. We then used \texttt{GALFIT} to fit these simulated galaxies. We noticed the underestimation of the flux by the \texttt{GALFIT} model in the simulated galaxies as well. We further noticed that a constant sigma map could recover the input parameters in simulated galaxies at this low count regime. Therefore, we used a constant sigma map for the analysis. The constant value was kept as the sky standard deviation (0.935 counts/pixel). Also, we had $\sim 34\%$ of the pixels with zero counts in the analysis region. As these zero values are mostly due to low background count (rather than bad pixels), we did not mask the zero values to fit the galaxy. The zero-point (ZP) magnitude provided in \citet{Tandon2020} is used to convert counts into magnitude units. We estimate the `point spread function (PSF)' from several point sources in the FOV using the \texttt{psf} task in the `Image Reduction and Analysis Facility (\texttt{IRAF})'\footnote{\url{https://iraf.net/}}. 

\begin{figure*}
\centering
	\includegraphics[width=13cm]{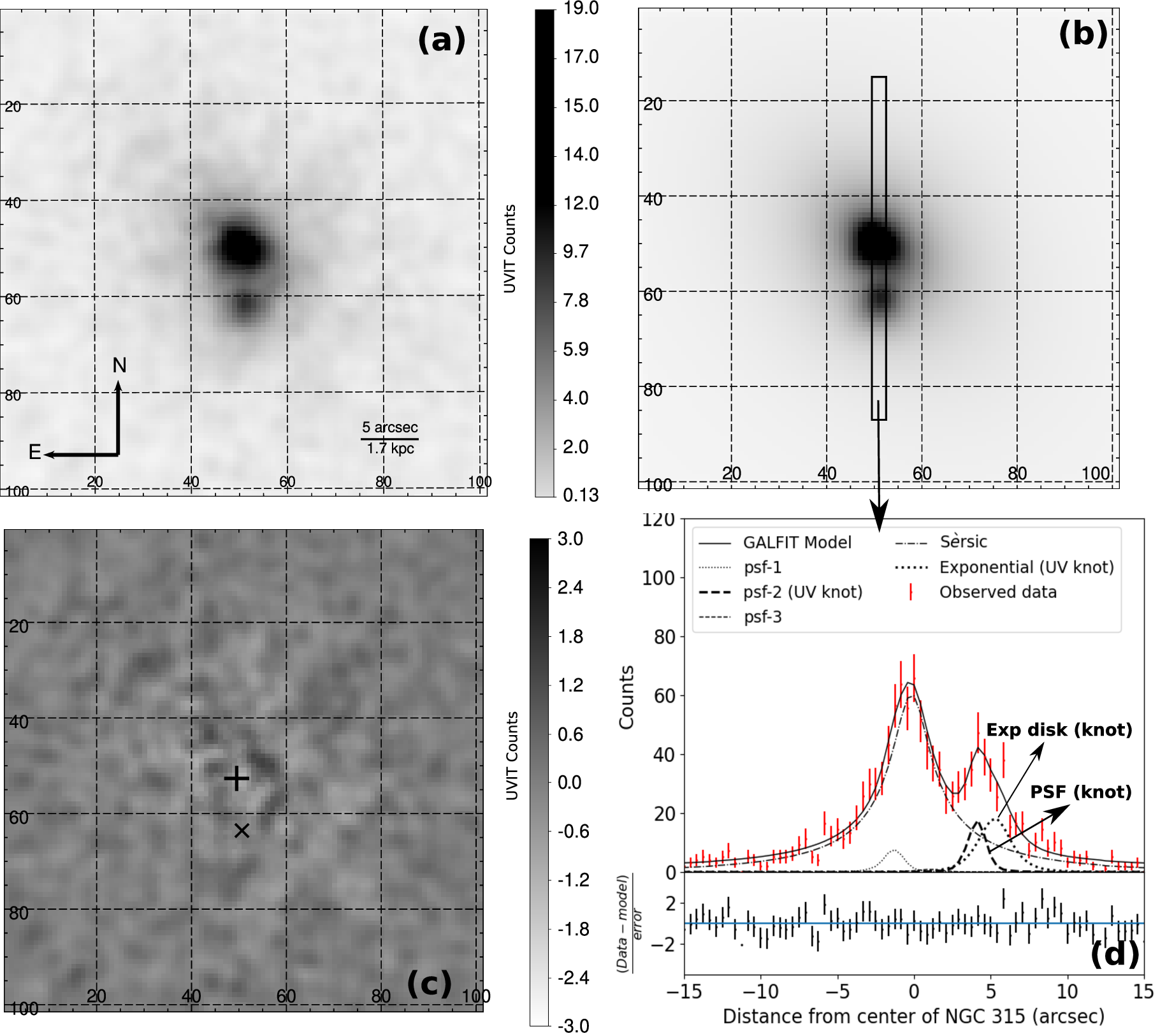}
    \caption{\texttt{GALFIT} modeling of NGC 315 in the central $100\times100$ pixels of UVIT F154W filter observations. Panel (a): Observed data. Panel (b): Best-fitting \texttt{GALFIT} model. Panel (c): 2D Residue image. The position of NGC 315 and the UV knot are shown in '$+$' and '$\times$' symbols, respectively. Panel (d): 1D profile of data and model components in a slice of 3 pixel width ($\sim$FWHM), along the box as shown in panel (b). For plotting the residual, a method similar to that in XSPEC \texttt{delchi} routine (\url{https://heasarc.gsfc.nasa.gov/xanadu/xspec/manual//node113.html}) is used.}
    \label{fig:galfit}
\end{figure*}

\begin{table*}
	\centering
	\caption{\texttt{GALFIT} fit parameters for UV data from UVIT. \label{Galfit_UV}}
	
	\begin{tabular}{lccccccc} % four columns, alignment for each
		\hline
		Component & $\Delta\alpha$ & $\Delta\delta$ &  &$r_e,_s$ &  &  &P.A.   \\
             & $(arcsec)$ & $(arcsec)$ &Mag  & $(arcsec)$ & $n$ &$q$  & (deg)  \\
        
             (1)& (2) & (3) & (4) & (5) & (6) & (7) & (8) \\
		\hline
            PSF  &$0.81$ & $1.24$ &$23.55 \pm 0.09$  & ...&... &...&... \\
            PSF &$1.49$ & $-0.91$ &$24.47\pm 0.20$  & ...&... &...&... \\
            S\'ersic & $0^a$& $0^a$&$18.56 \pm 0.04$ &$12.47\pm 0.74$ & $2.77 \pm 0.11$ & $0.76 \pm 0.01$ & $-42.01 \pm 2.43$ \\
            {\bf PSF (Knot)} &$0.22$ & $-4.3$ & $23.49 \pm 0.13$ & ...& ...&...&... \\
            {\bf Exp disc (Knot)} & $0.02$&$-5.44$ & $22.18 \pm 0.05$ &$0.88 \pm 0.06$ & ... &$0.70 \pm 0.06$ &$74.33 \pm 13.61$ \\
            Sky & ... & ... & $0.522^{b}$ & ... & ... & ...& ...\\
            %sky &... & ...& $0.695 \pm 0.003$ counts/pixel &... &... &... \\
            \hline
	\end{tabular}

  {\raggedright  \textsc{Notes.--} Col. (1): \texttt{GALFIT} component. Col. (2), Col. (3): Offset from the galaxy's centre. Col. (4): AB magnitude of the component except for `Sky' for which it is in the unit of counts/pixel. Col. (5): The effective radius ($r_e$) of the S\'ersic law for the  `S\'ersic' component; the scale length ($r_s$) of the exponential disc for the `Exp disc' component. Col. (6): The S\'ersic exponent 1/n. Col. (7): Axis ratio. Col. (8): Position angle. \\$^{a}$ The value was kept fixed during the fit. $^{b}$ For Sky component Col. (4) is in counts/pixel unit and kept fixed during the fit. \par}

\end{table*}

We initially fitted the galaxy by masking the UVIT knot. We utilised the method provided \citet{Ananthamoorthy2024} to detect and mask the UV knot. The pixels corresponding to the knots are identified using the segmentation map from \texttt{SExtractor} \citep{Bertin1996}.  We used the `psf+S\'ersic+sky' model for fitting the galaxy. The central position of the S\'ersic component of the model was fixed to the centre of the galaxy during the fit. Then, we unmasked the pixels corresponding to the UV knots. The UV knot could be modelled with additional `psf' and `expdisc' components. An additional small residue is observed in the central region, which is accounted for by an additional `psf' component. The further addition of the component did not improve the residual. Smoothed data, model, and residue maps are provided in Fig.~\ref{fig:galfit}. Panel (d) of Fig.~\ref{fig:galfit} shows the 1D profile of data and model components in a slice of 3 pixel width ($\sim$FWHM), along the box as shown in panel (b). The 1D radial profile clearly shows the enhanced UV emission corresponding to the UV knot at $\sim5^{\prime\prime}$ away from the centre. The best-fitting model parameters are provided in Table~\ref{Galfit_UV}. We noticed that the offset between the centre of the `psf' component and that of the `S\'ersic' component is $\sim 1.5^{\prime\prime}$. This offset value is in agreement with the PSF ($\sim 1.3^{\prime\prime}$) of the UVIT. The knot of UV emission, which has been accounted for with the `psf+exponential disc' profile, has a AB magnitude of $21.89\pm0.05$, i.e.,  flux of $8.0\pm0.33 \times 10^{-17}$ erg cm$^{-2}$ s$^{-1}$ \AA$^{-1}$ in the FUV F154W filter \citep{Tandon2017, Tandon2020}. We also calculated the ratio of the UV flux of the knot to the underlying smooth emission in the aperture of $\sim 1.5^{\prime\prime}$ centred at the UV knot using the \texttt{GALFIT} model components. The UV knot has a $\sim 2$ factor higher flux than the underlying continuum emission. The second central psf component is not statistically very significant ($\sim 1.5\sigma$). Due to its low significance, we do not assign a physical interpretation. It is to be noted that though the central region of the galaxy is detected in H$\alpha$ emission, no significant H$\alpha$ emission is observed along the direction of the UV knot as presented in fig. 6 of \citet{Masegosa2011}. Also, recent studies by \citet{Hermosa2024} report that H$\alpha$ emission is detected only up to the central $4^{\prime\prime}$($\sim1.4$ kpc) region of the galaxy.

\subsection{SFR of the UV knot } \label{sec:SFR}
We derived the SFR of the UV knot using the UV magnitude estimated from the \texttt{GALFIT} modelling. The UV magnitude is corrected for galactic extinction using the relation by \citet{Cardeli1989} considering A$_V$ = 0.177 \citep[NED;][]{Schlafly2011}. We utilised UV colour to derive the intrinsic extinction corrected SFR as outlined below \citep{Rubinur2024}.

The {\sl GALEX} angular resolution is $\sim$ three times lower than that of UVIT and $\sim$ two times lower that of UVOT. As a result, only a fraction of the knot flux is captured in small apertures for {\sl GALEX}, while larger apertures suffer strong contamination from the main body UV emission. In addition, the photometric uncertainties of {\sl GALEX} fluxes are not better than those from UVIT and UVOT. Therefore, we have considered only UVIT and UVOT observations to derive the NUV-FUV colour and for further analysis. As the resolutions of the UVIT and UVOT images are different, the UVIT image is convolved with the Gaussian kernel of FWHM $\sim 2.84^{\prime\prime}$ to simulate the UVOT angular resolution. The background contribution external to the galaxy is calculated in a few source-free regions (size of 5$^{\prime\prime}$ each) away from the galaxy. For each filter, the median of the background per pixel in these source-free regions is subtracted. We calculated the flux in 1.5$^{\prime\prime}$ (diameter of $\sim$ FWHM of the UVOT observation) centred around the UV knot in each filter. As the {\it uvm2} filter significantly overlaps with the 2175 $\mbox{\AA}$ dust feature, we did not consider it to derive the UV spectral slope. Further, the UV emission can have a contribution from the AGN and the underlying old stellar population. The central PSF component (likely due to AGN) of the UVIT \texttt{GALFIT} model suggests that the AGN contribution is negligible ($\leq 0.3\%$) in the region used for colour calculation. Therefore, we have ignored the AGN contamination in estimating the colour. However, the S\'ersic component has a significant contribution ($\sim 42\%$ after smoothing to UVOT resolution) in this region. We used \texttt{Starburst99} \citep{Leitherer1999} models, as discussed below, to subtract the underlying galaxy emission, assuming this underlying UV emission is due to the main stellar population of the galaxy.

\begin{table*}
	\centering
	\caption{UV magnitude/flux in an aperture of $1.5^{\prime\prime}$ radius centred at UV knot for the filters used in this work \label{Knot_flux}}
	
	\begin{tabular}{lcccccc} % four columns, alignment for each
		\hline
		Instrument & Filter & Effective &Bandwidth& Observed & AB Magnitude  & Flux \\
             &  & $\lambda(\text{\AA})$& $\Delta\lambda(\text{\AA})$&AB Magnitude &corrected for  &corrected for   \\
            & & &&& old stellar  & old stellar population   \\
            & & &&& population&($\times 10^{-17}$erg cm$^{-2}$ s$^{-1}$\AA$^{-1}$)\\
		\hline
            {\sl AstroSat}-UVIT  &F154W & $1541^a$  &$380^a$ &$22.07\pm0.07$&$22.61\pm0.12$ & $4.1\pm0.5$ \\
            {\sl Swift}-UVOT     &uvw2 & $1991^b$   &$657^c$ &$21.50\pm0.06$&$21.74\pm0.08$ &$5.5\pm0.4$  \\
                                & uvw1& $2486^b$    &$693^c$&$20.70\pm0.06$&$20.82\pm0.06$  &$8.3\pm0.5$   \\
            \hline
	\end{tabular}

  {\raggedright  \textsc{Notes.--} $^a$ \citet{Tandon2017}, $^b$ \url{https://heasarc.gsfc.nasa.gov/docs/heasarc/caldb/swift/docs/uvot/uvot_caldb_AB_10wa.pdf}, \\$^c$ \citet{Poole2008}. \par}

\end{table*}

The age of the underlying stellar population was considered to be $5.4\pm1.5$ Gyr \citep{Trager2000}. We considered a continuous SF model with a global SFR value of $0.255$ M$_\odot$yr$^{-1}$ as estimated by \citet{Kolokythas2022} for NGC 315. We generate simulated evolutionary tracks considering the Salpeter initial mass function \citep[IMF;][]{Salpeter1955}, and the Padova evolutionary tracks \citep[][and references therein]{Leitherer1999, Vazquez2005}. \citet{Zhang2008} estimated the metallicity (Z) for NGC 315 to be 1.29 times the solar metallicity (0.014), resulting in Z$\sim0.018$ for NGC 315. We considered the metallicity of Z = 0.020, the closest available metallicity in \texttt{Starburst99} to simulate stellar evolutionary tracks. The synthetic spectra are then redshifted and convolved with the filter responses of UVIT \citep{Tandon2020} and UVOT\footnote{\url{https://heasarc.gsfc.nasa.gov/docs/heasarc/caldb/swift/docs/uvot/}} to obtain the expected flux in the UVIT and UVOT filters\footnote{Filter responses for both UVIT and UVOT are obtained from \url{http://svo2.cab.inta-csic.es/theory/fps/}}. The UVIT flux from the S\'ersic component (after smoothing to UVOT resolution) in the region of colour calculation was used to rescale the UVOT filter fluxes to obtain the contribution from the underlying stellar population in the UVOT filters. These fluxes are subtracted from the observed flux in UVOT filters to obtain the flux corresponding to the UV knot. We verified that the correction for the underlying old stellar population is largely insensitive to the choice of IMF and SFR. This is because we use only the \texttt{Starburst99} spectral template, not the normalisation. We then rescale this template to match the flux from the UVIT S\'ersic profile in the region of colour calculation. For instance, adopting a Kroupa IMF \citep{Kroupa2001} or varying the SFR (e.g., 1 M$_\odot$ yr$^{-1}$) changes the relative flux between UVOT and UVIT filters by only $\sim 1\%$. The observed AB magnitude, magnitudes after subtracting the contribution from the old stellar population in a $3^{\prime \prime}$ aperture diameter centred at the knot from UVIT and UVOT observations, is provided in Table~\ref{Knot_flux}.

We used equation~(4) of \citet{Nordon2013} to derive the UV spectral slope, $\beta$ (flux $\propto$ $\lambda^{\beta}$). We fitted a straight line to the magnitude vs. $-2.5\log_{10}(\lambda)$ in UVIT F154W, UVOT {\it uvw1}, and {\it uvw2} filters (after correcting for underlying galaxy emission), which yielded $\beta= 1.46\pm0.22$. The calculated $\beta$ is used to derive the color excess $E(B-V)$, using equation~(1a) of \citet{Reddy2018}. The estimated $E(B-V)$ is $0.87\pm0.046$ mag. Further, equations~(2),~(3), and ~(4) of \citet{Calzetti2000} are used to derive the FUV attenuation. Using the relative visual extinction ($R'_V$) of $4.05\pm0.8$, we calculated the FUV attenuation (A$_{\text{FUV}}$) as $3.90\pm0.45$ mag. The attenuation-corrected FUV flux is converted to SFR using the relation by \citet{Iglesias2006}. The derived SFR, averaged over 100 Myr, for the UV knot is $0.23\pm0.10$ $M_{\odot}$ yr$^{-1}$. The derived SFR surface density ($\Sigma_{\text{SFR}})$ is $0.818\pm0.36$M$_\odot$yr$^{-1}$ kpc$^{-2}$. We considered the scale length of the exponential profile (dominant component for the UV knot) from \texttt{GALFIT} as the radius for the UV knot for deriving the $\Sigma_{\text{SFR}}$.

\begin{figure}
\centering
	\includegraphics[width=9cm]{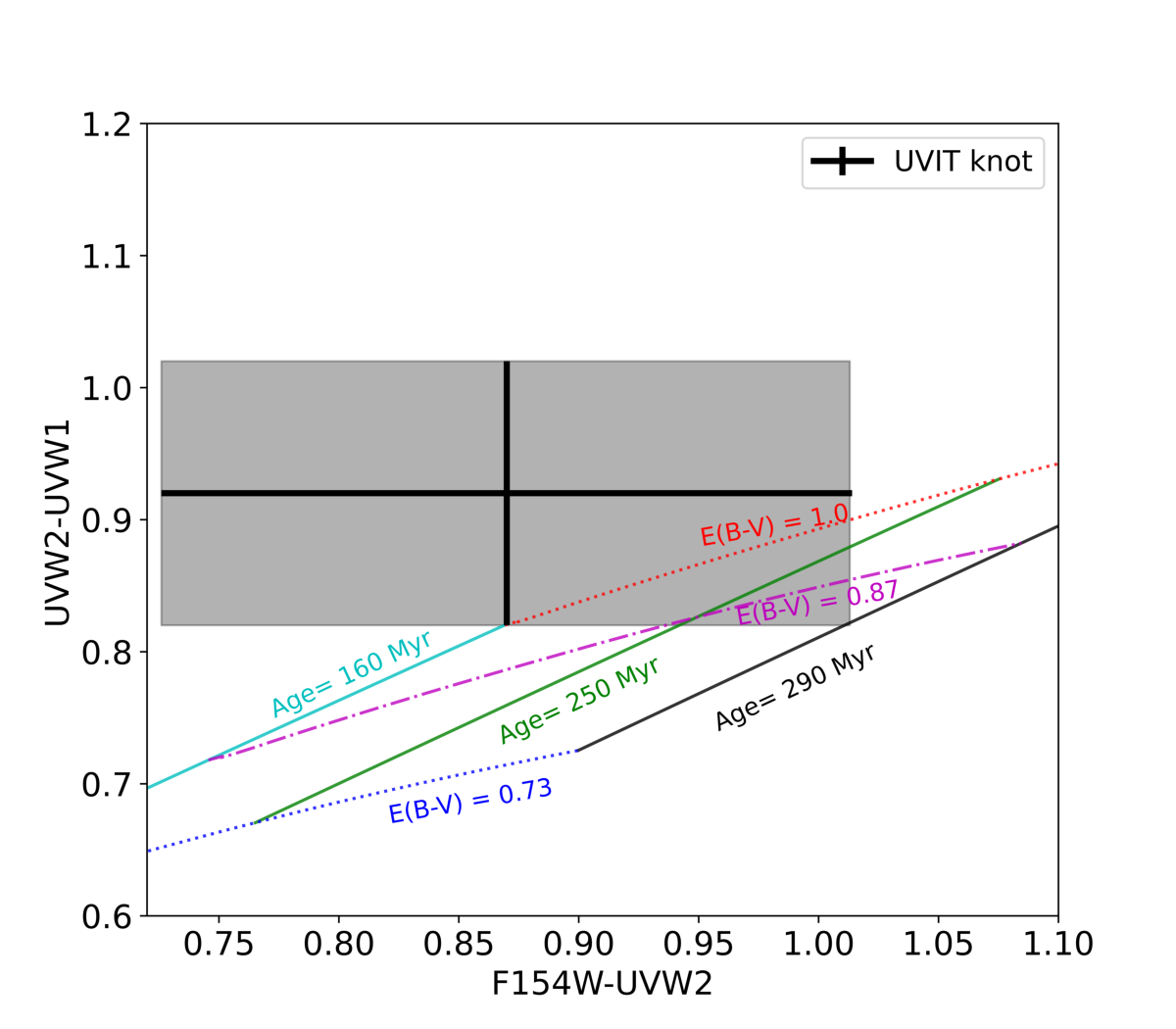}
    \caption{Colour-colour diagram for UV knot overlaid on the stellar population synthesis tracks and colour excess, E(B-V), for different ages.}
    \label{fig:cmd}
\end{figure}

\subsection{Age and mass of star-forming knot}\label{Age_Mass}
The age and stellar mass of the star-forming knot are derived using \texttt{Starburst99} models \citep{Leitherer1999}. We have subtracted the contribution from the old stellar contribution (as discussed in section~\ref{sec:SFR}). Based on H$\alpha$ emission and UV-IR colours (discussed in detail in section~\ref{sec:UV_main_body}), we have considered that the continuous SF in the main body of the galaxy is minimal. Hence, a single-burst SF model is adopted to model the UV knot emission. Except for the single-burst SF assumption, all other model parameters are similar to as described in Section~\ref{sec:SFR}.

Since the colours are independent of the stellar population mass, a colour–colour diagram is used to estimate the age and intrinsic attenuation of the knot. The \texttt{Starburst99} model spectra are reddened using \citet{Calzetti2000} attenuation law considering various E(B-V) values. The model spectra are then redshifted and convolved with the respective filter transmission curves for comparison with the observed colours.

We notice that if we simultaneously fit the age and extinction, the lower limit on age and the upper limit on $E(B-V)$ cannot be constrained. However, as the $E(B-V)$ derived from UV colours and optical colours (discussed in section 3.5) are consistent with each other, we put a constraint on the value of $E(B-V)$ during the fit. Considering the $3\sigma$ error on $E(B-V)$, we varied it from $0.73$ to $1.0$ mag to estimate the age. Within this $E(B-V)$ range, the age of $160-290$ Myr is consistent with observation. It is a young to intermediate stellar population considering the main stellar population of the galaxy \citep[$5.4\pm1.5$ Gyr;][]{Trager2000}. The position of the UV knot in the colour-colour plot is provided in Fig.~\ref{fig:cmd} along with stellar population evolutionary tracks for different ages and $E(B-V)$ values.

Considering the age of $200$ Myr for the stellar population and $E(B-V)$ of $0.87$ mag, we derive the stellar mass of the single-burst stellar population using the FUV magnitude. The mass was estimated to be  $2.3^{+0.9}_{-0.65} \times 10^8$ M$\odot$. We also noticed that considering Kroupa IMF \citep{Kroupa2001} would also yield similar results for these parameters, indicating no significant dependence of the derived parameter on the choice of IMF.

\subsection{Dust distribution}\label{dust_dist}

\begin{figure*}
\centering
	%\includegraphics[width=14cm]{HST_galfit_April_2025.eps}
    \includegraphics[width=13cm]{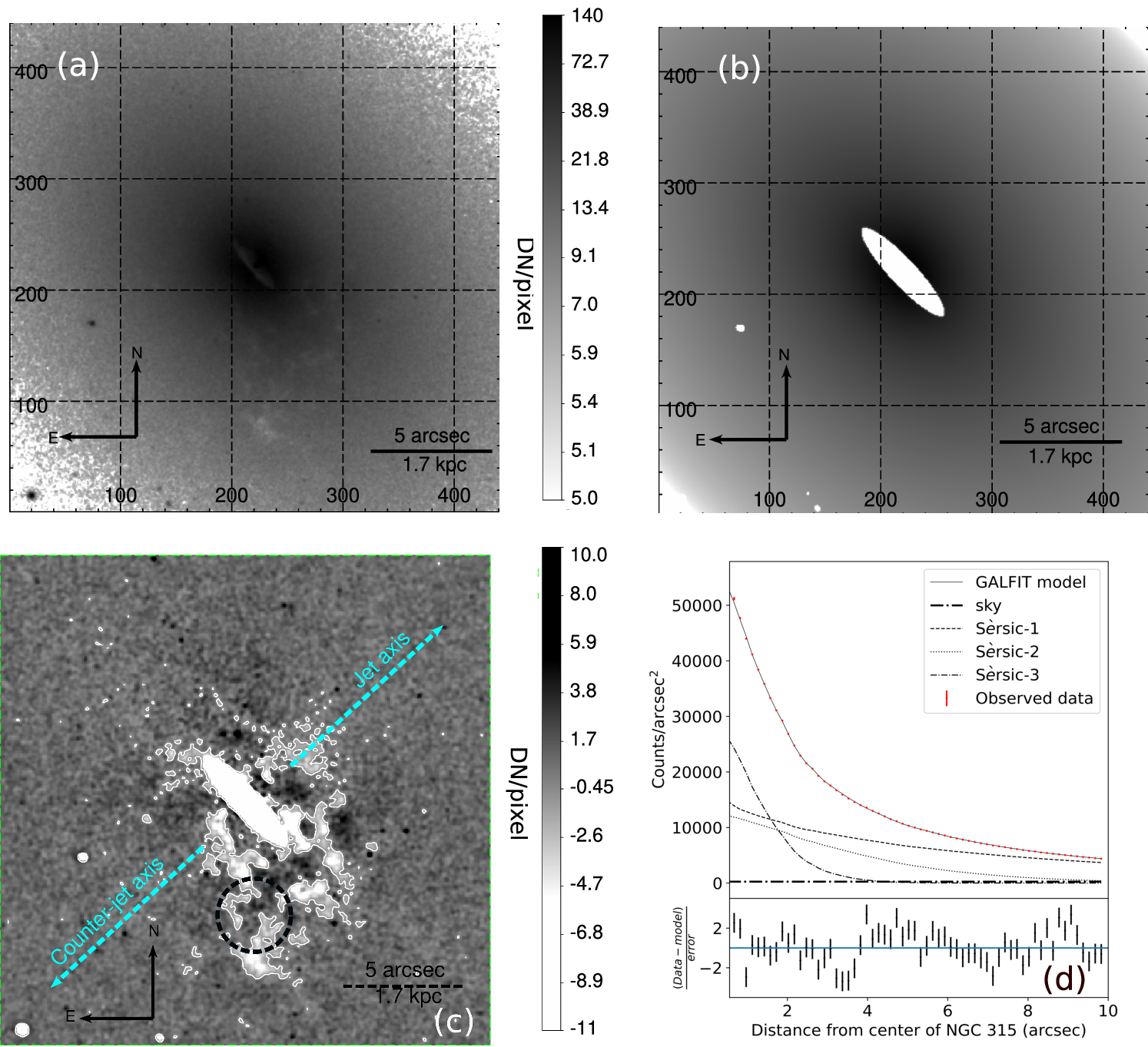}
    \caption{\texttt{GALFIT} modeling of NGC 315 in the central $440 \times 440$ pixels of HST F555W filter observations. Panel (a): Observed data. Panel (b): Best-fitting \texttt{GALFIT} model. The white region corresponds to the central dust disc region, which is masked during the fit. Panel (c): 2D Residue image. White contours correspond to dust patches detected in absorption. The dashed circle corresponds to 1.5$^{\prime\prime}$ centred around the UVIT knot. Lower right panel: 1D radial profile, model components, and residue. For plotting the residual, a method similar to that in XSPEC \texttt{delchi} routine is used.}
    \label{fig:galfit_hst}
\end{figure*}

\begin{table*}
	%\centering
	\caption{\texttt{GALFIT} fit parameters for optical data from optical F555W filter observations. \label{Galfit_optical}}
	
	\begin{tabular}{lccccccc} % four columns, alignment for each
		\hline
		Component & $\Delta\alpha$ & $\Delta\delta$ &  &$r_e$ &  &  &P.A.   \\
             & $(arcsec)$ & $(arcsec)$ &Mag  & $(arcsec)$ & $n$ &$q$  & (deg)  \\
        
             (1)& (2) & (3) & (4) & (5) & (6) & (7) & (8) \\
		\hline
            S\'ersic-1 & $-0.16$& $0.0$&$10.60 \pm 0.04$ &$20.17\pm 0.71$ & $1.39 \pm 0.08$ & $0.70 \pm 0.01$ & $43.28 \pm 0.09$ \\
            S\'ersic-2 & $0.14$ & $0.13$ &$12.93\pm0.11$ &$4.61\pm 0.06$ & $0.71\pm0.02$ & $0.72 \pm 0.01$ & $45.39 \pm 0.14$ \\
            S\'ersic-3 & $0.0$ & $0.0$ &$14.07\pm0.04$ &$1.55\pm 0.01$ & $0.68\pm0.01$ & $0.79 \pm 0.01$ & $36.37 \pm 0.35$ \\
            Sky  & ... & ... & $0.66^{a}$ & ... & ... & ... & ...\\
            \hline
	\end{tabular}

    {\raggedright  \textsc{Notes.--} Col. (1): \texttt{GALFIT} component. Col. (2), Col. (3): Offset from the galaxy's centre. Col. (4): AB Magnitude of the component. Col. (5): The effective radius ($r_e$) of the S\'ersic law for the `S\'ersic' component. Col. (6): The S\'ersic exponent (n). Col. (7): Axis ratio. Col. (8): Position angle. \\$^{a}$ For Sky component Col. (4) is in DN/pixel unit, and it is kept fixed during the fit  \par}
\end{table*}
We utilised high-resolution {\sl HST} images to study the galaxy's dust distribution. {\sl HST} observations have revealed a central dust disc and several patches of dust in the central $10^{\prime\prime}$ region of the galaxy \citep{Capetti2000, Kleijn1999}. To understand the distribution of these irregular dust patches, we used \texttt{GALFIT} \citep{Peng2002, Peng2010} to fit the galaxy emission profile. We utilized central $441 \times 441$ pixel regions ($\sim 20^{\prime \prime} \times 20^{\prime \prime}$) of F555W filter observations from the WFPC2 instrument. The PSF is obtained from the Hubble archive\footnote{\url{https://mast.stsci.edu/portal/Mashup/Clients/Mast/Portal.html}} near the observing time of the galaxy. The galaxy's central region, containing the dust disc, is masked during the fit. To mask the central dust disc, we considered twice the size of the dust disc values provided in Table~4 of \citet{Kleijn1999}. Therefore, an elliptical region of $2.5^{\prime\prime}$ as the semi-major axis with an axis ratio (b/a) of 0.23 and a position angle of $40^{\circ}$ centred on NGC 315 was used to mask the central dust disc. We also mask the other point sources detected in the analysis region using the segmentation map from \texttt{SExtractor} \citep{Bertin1996}. 

We used \texttt{GALFIT} generated sigma map for the analysis. However, for NGC 315, the galaxy occupies most of the region of the observation. Therefore, the background calculated from the image will be overestimated. To overcome this issue, we calculated the sky value and the sky root mean square (RMS) values as given below.

For sky background, we considered the background value of 0.01 e$^-$/sec/pixel, which is the expected background level in the F555W filter as provided in the WFPC2 instrument handbook\footnote{\url{https://www.stsci.edu/instruments/wfpc2/Wfpc2_hand/HTML/W2_1.html}}. The sky RMS value is calculated by adding the contribution due to the sky background (Poisson noise), readout noise (5.3 e$^-$), and a dark current (0.0045 e$^-$/sec/pixel), in quadrature \citep{Bonfini2014}. The background RMS value is found to be $\sim$ 6.1 e$^-$/pixel. These sky background and sky RMS values are provided as input to \texttt{GALFIT} to generate the sigma map.

We used three S\'ersic profiles to model the galaxy. The residual map revealed irregular dusty filaments near the central region of the galaxy, which have negative counts due to absorption. As these dusty filaments can affect the galaxy fitting, we identified and masked these pixels to fit the galaxy again. The pixels associated with the dusty filaments are calculated as follows. We calculated the local fluctuations ($\sigma$) in the residual image near these dust filaments. Pixels with counts $3\sigma$ lower than the mean counts ($\sim$zero in residual image) were identified as associated with dust filaments. The galaxy is fitted again after masking these dust filaments. The parameters of the fitted model (red.$\chi^2$ = $119315/186167 = 0.641$) are provided in Table~\ref{Galfit_optical}. Smoothed data, model, residue, and 1D radial profiles are provided in Fig.~\ref{fig:galfit_hst}.  

The residual image provided in panel (c) of Fig.~\ref{fig:galfit_hst} reveals the dust distribution in the galaxy. The knot of UV emission observed in UVIT appears to reside in the dust filaments connected to the southeast direction.

\subsection{Optical dust extinction and dust mass:} \label{sec:dust_mass}
We utilised {\sl HST} F555W and F814W filter observations to derive the optical extinction and dust mass of the filamentary structure. The AB magnitude in each filter is calculated in an aperture of 3$^{\prime\prime}$ centred at the position of the UV knot. The derived colour (F555W-F814W mag) was 0.902 in the AB magnitude system. It is further corrected for galactic extinction \citep{Cardeli1989} considering A$_V$ = 0.177 \citep[NED;][]{Schlafly2011}. This observed colour in the optical band is due to the intrinsic emission of the stellar population in the galaxy and absorption due to the presence of dust intrinsic to the galaxy \citep{Mathis1990}. The UV knot has no optical counterpart; therefore, the optical emission is expected to be dominated by the underlying old stellar population of age $5.4\pm1.5$ Gyr \citep{Trager2000}. We used \texttt{Starburst99} \citep{Leitherer1999} to estimate the intrinsic emission of this stellar population. The parameter/models used are similar to as described in Section~\ref{sec:SFR}. The expected magnitude in each filter is derived by convolving the redshifted simulated spectra with the corresponding filter responses\footnote{\url{http://svo2.cab.inta-csic.es/theory/fps/}}. The expected colour from the stellar population of $5.4\pm1.5$ Gyr was found to be $0.28\pm0.04$, yielding a colour excess, E(F555W-F814W), of $0.62\pm0.04$ mag. For the \citet{Calzetti2000} dust attenuation law, this color excess corresponds to A$_{\text{V}}$ (at 5500 $\text{\AA}$) of $1.65\pm0.34$ mag. Further, this colour excess corresponds to A$_{FUV}$ of $4.15\pm0.42$ mag. It is in good agreement with that derived from the UV spectral slope ($3.95\pm0.45$ mag). However, it should be noted that the intrinsic colour derived for the stellar population is sensitive to the SF history of the galaxy.

We also calculated the dust mass utilising the visual extinction \citep[$A_V$;][]{Dokkum1995}. The dust mass (M$_{dust}$) is calculated as M$_{dust}$ = $\Sigma\Gamma_V^{-1}A_V$, where $\Sigma$ is the dust feature area in kpc$^{2}$, $\Gamma_V$ is the visual mass absorption coefficient, and $A_V$ is the optical extinction. As the differences between the Galactic and elliptical galaxy extinction curves are generally small \citep{Goudfrooij1994}, we adopted the Galactic value for the visual mass absorption coefficient $\Gamma_V$ = $6\times 10^{-6}$ mag kpc$^2$ M$_{\odot}^{-1}$. The pixels associated with dust filaments are identified using the residual image, as explained earlier in this section, which are further used to derive the area corresponding to the dust filaments. 
The estimated $A_V$ for the pixels associated with filamentary dust structure was  $1.72\pm0.36$ mag, which is in very good agreement with that calculated at the position of the UV knot. The dust mass obtained for these filamentary dust structures was $5.9 \pm 1.2 \times 10^{5}$ M$_{\odot}$. The area covered by the dust patches was $2.07$ kpc$^2$ yielding a dust surface density ($\Sigma_{\text{dust}}$) of $\sim 2.9\pm0.6\times10^{5}$ M$_\odot$ kpc$^{-2}$. We also separately estimated the $A_V$ in a few circular apertures within the central dust disc to estimate its dust mass. The estimated $A_V$ was $2.0\pm0.4$ mag and the dust mass was $\sim 1.1\pm0.2\times10^{5}$ M$_\odot$ for the central dust disc.

%%%%%%%%%%%%%%%%%%%%%%%%%%%%%%%%%Discussion%%%%%%%%%%%%%%

\section{Discussions and Conclusions}
%\subsection{Origin of UV emission:}

\subsection{Origin of UV knot emission}
The resolved UV knot emission from UVIT observation and the different UV colours in the central region and the knot indicate a different origin of the UV knot emission compared to the nuclear (AGN) emission. No significant H$\alpha$ emission is observed in the direction of the UV knot \citep[fig.~6 of ][]{Masegosa2011}. The non-detection of the H$\alpha$ emission suggests that the UV emission is unlikely due to any ionization, including the AGN.

The UV emission due to old and metal-rich stellar populations could be responsible for the `UV upturn' in many elliptical galaxies \citep[e.g.,][]{Yi2008, Kaviraj2007}. However, such UV emission is expected to have a smooth distribution following the underlying older stellar population \citep{Pandey2024}. Moreover, studies have suggested the distinct knot of UV emission in elliptical galaxies and/or BCGs could be due to recent SF in the galaxy \citep{Koekmoer1999, Odea2004, Odea2010, Donahue2015, Pandey2024}. Therefore, UV knot emission, which is a factor of two higher than the underlying smooth UV emission, is likely the result of ongoing SF within the galaxy.

The UV emission from the knot region is heavily obscured by dust with A$_{\text{FUV}}$ of $3.90\pm0.45$ mag derived using UV colour and $4.15\pm0.42$ mag using optical colours. Possible significant obscuration of the FUV emission in BCG is also reported by other studies \citep[e.g.,][]{Odea2010}. The derived UV SFR, averaged over 100 Myr, is $0.23\pm0.10$ M$_{\odot}$yr$^{-1}$ for UV knot. This value is considerably higher than what is typically observed in elliptical galaxies \citep[$10^{-4}$ M$_{\odot}$yr$^{-1}$;][]{Ford2013}, suggesting active SF in the UV knot. The age of the stellar population obtained is between $\sim 160-290$ Myr from the UV colours, suggesting an intermediate stellar population. Absence of significant H$\alpha$ emission also indicates that the age of the stellar population is older than $\sim 20$ Myr \citep{Kennicutt1998}. Furthermore, the UV-emitting knot has no optical counterpart, and given the current resolution of the IR observations ({\it Sptizer}), it is difficult to draw definitive conclusions about the presence of excess IR emission around the knot. Also, future observations are essential to spectroscopically confirm the redshift of the UV knot.

\subsection{UV emission from the main body of the galaxy}\label{sec:UV_main_body}
\citet{Kolokythas2022} estimated the UV SFR (not corrected for attenuation) to be 0.255 M$_\odot$ yr$^{-1}$, assuming that the entire UV emission is associated with the SF. Using the WISE W3-W4 band colour, they suggest possible dust-obscured SF in NGC 315. However, the FUV-K \citep[$9.6\pm 0.13$;][]{Gil2007}, NUV-K \citep[$8.61\pm 0.05$;][]{Gil2007, Hoffer2012} and J-K \citep[$1.0\pm0.03$;][]{Gil2007, Donahue2015} colours of NGC 315 indicate that it could be a quiescent BCG dominated by an old stellar population. Recent observations have also reported that H$\alpha$ emission in NGC 315 is contained within a central $4^{\prime \prime}$ \citep{Hermosa2024}. This further suggests that the bulk of the UV emission from the main body of the galaxy may not be due to recent SF in the galaxy. Consequently, in this work, we have considered that the main body UV emission is dominated by the old stellar population of age $5.4\pm1.4$ Gyr \citep{Trager2000}. A reliable determination of the intrinsic SFR for the main body would require future detailed modelling of the galaxy emission utilising multiband photometric and/or spectroscopic observations. 

It is also worth noting that the age of the main stellar population ($5.4\pm1.4$) is relatively young for massive BCGs or elliptical galaxies \citep[Stellar mass $1.17\times10^{12}$ M$_\odot$;][]{Kolokythas2022, Gozaliasl2024, Kuntschner2010}. A relatively younger dominant stellar population could be due to SF triggered by mergers/interactions with neighbouring systems in its early evolution stage.

\subsection{UV emission from the AGN}\label{sec:UV_AGN}
The central PSF/AGN component in the FUV (section~\ref{UV_knot}) is $\sim$ hundred times fainter than the galaxy (S\'ersic component; Table~\ref{Galfit_UV}). The estimated A$_V$ of $\sim 2$ mag for the central dust disc (in section~\ref{sec:dust_mass}) corresponds to A$_{\text{FUV}}$ of $\sim 5$ mag, about one mag higher than at the UV knot. This value indicates that UV emission from the central region is severely attenuated. Studies on nearby LINER-type (Similar to NGC 315) galaxies have also suggested obscuration by dust as a primary reason for non-detection/weak UV emission in LINER \citep{Pogge2000}. Further, the intrinsic FUV emission of the AGN may be low, as suggested by broadband SED modelling of the NGC 315 core region \citep{Tomar2021}, possibly due to the FUV being at the falling edge of Synchrotron emission. Therefore, both local extinction and intrinsically low UV emission could contribute to the lower FUV emission of the central PSF component compared to the galaxy.

\subsection{Gas and dust density at the position of UV knot}
We noticed that no significant CO(2-1) emission is observed with the `Atacama Large Millimeter/submillimeter Array (ALMA)' observation \citep{Boizelle2021} along the UV knot or any other optical dust patches. Therefore, we used the dust-to-gas ratio (DGR) to constrain the gas density and examine possible reasons for the absence of CO emission. 

We consider a DGR of $10^{-3}$, which is consistent with that observed for elliptical galaxies \citep[$10^{-2}-10^{-3}$;][]{Davis2015}, and $\Sigma_{\text{dust}}$ derived using A$_V$ (Section~\ref{dust_dist}), which yield $\Sigma_{\text{gas}}$ of $\sim 290\pm60$ M$_\odot$pc$^{-2}$. This value of $\Sigma_{\text{gas}}$ results in a molecular gas number density ($n$) of $\sim 30-46$ cm$^{-3}$ for the gas clouds, considering the resolution of ALMA observation ($\sim 0.3^{\prime\prime}$; cloud of $\sim 100$ pc scale). It is well below the critical density ($n_{\text{crit}} \sim 2 \times 10^{4}$ cm$^{-3}$) for CO(2-1) emission \citep{Krumholz2017, Schoier2005}. However, the optical depth ($\chi$) of the medium can effectively reduce $n_{\text{crit}}$. We used the relation provided by \citet{Narayanan2014} to derive the effective critical density ($n_{\text{crit, eff}}$). Equation~(18) of \citet{Narayanan2014}, provides the relation between $n_{\text{crit, eff}}$ and $n_{\text{crit}}$ in the form $n_{\text{crit, eff}} = kn_{\text{crit}}$, where $k=\frac{1}{1+(3/8)\chi}$. The value of $\chi$ can range over $\sim10-50$ (CO(1-0) emission) for the observed $\Sigma_{\text{SFR}}$, as provided in fig.~5 of \citet{Narayanan2014}. Even considering an higher value of 50 for $\chi$ yields $n_{\text{crit, eff}}\simeq 10^{3}$ cm$^{-3}$. Despite this reduction, $n$ is still much lower (factor of $\sim10$) than $n_{\text{crit, eff}}$. Consequently, a possible reason for the lower CO emission could be the low gas density ($n < n_{\text{crit}}$).

If we consider a Kennicutt-Schmidt (KS) relation of the spatially resolved SF region \citep[equation~(9) in][]{Leroy2013}, yields $\Sigma_{\text{gas}}$, of $\sim 0.74-8.0 \times 10^{3}$ M$_\odot$pc$^{-2}$, after inclusion of the error in $\Sigma_{\text{SFR}}$ and a scatter of $0.3$ dex in the KS relation. This value is much higher than that derived using DGR and is comparable to $\Sigma_{\text{gas}}$ of the central molecular disc \citep[peak density of $3.1\times10^{3}$M$_\odot$pc$^{-2}$;][]{Boizelle2021} which is detected in ALMA observation. Therefore, the absence of significant CO emission along the UV knot suggests that the actual $\Sigma_{\text{gas}}$ could be lower than that from the KS relation. The lower $\Sigma_{\text{gas}}$ than the value calculated from the KS relation also indicates that the gas at UV knot could have a lower depletion time scale ($\tau_\text{dep}$) or higher SF efficiency (inverse of $\tau_\text{dep}$), placing it above the standard KS relationship. The absence of CO emission along the dust feature is also observed in the southern dust lane of NGC 3100 \citep{Ruffa2022}, which also hosts a radio AGN, PKS 0958-314. A more sensitive and detailed study of such gas clouds is essential to further comment on the nature of gas in these regions.

\subsection{Possible mechanism for ongoing SF}

Potential mechanisms for ongoing SF at the UV knot include triggered SF due to AGN feedback or SF in the gas supplied due to merger events. Each of them is discussed below.

{\bf (a) Role of mergers:} No morphological disturbances are observed in NGC 315, suggesting the absence of recent major merger events. Moreover, \citet{Morganti2009} noticed the lack of an extended disc/ring-like structure in the HI emission, usually observed in galaxies with a recent history of major mergers. Based on this observation, they suggested that the galaxy would not have undergone any major merger in the recent past. However, they also detected five galaxies with HI emission in the NGC 315 group, upon which they inferred that the NGC 315 group is a gas-rich system, and many gas clouds would have fallen into the galaxy, supplying gas. HST observations reveal that the UV knot resides in the projected dust filaments connected to the southeast direction.

\citet{Kaviraj2014} and \citet{Kaviraj2011} have suggested that at least $14\%$ of the SF in elliptical galaxies could be due to minor mergers. Also, typically, the residual ongoing SF in elliptical galaxies has often been attributed to minor merger-driven SF \citep[e.g.,][]{Pandey2024}. Therefore, one of the possible scenarios for triggered SF in NGC 315 could be minor mergers. However, it is also to be noted that no significant clumpy UV emission is observed in other regions along the dust filaments.
Therefore, if the UV SF corresponds to the gas clouds associated with dust filaments, an additional mechanism could also be playing a role in triggering the SF at the position of the UV knot.

{\bf (b) Role of AGN feedback:} NGC 315 contains a FR-I radio jet \citep[e.g.,][]{Liang2006} with derived jet power ranges between $\sim 10^{43-44}$ erg s$^{-1}$ in the literature \citep[e.g.,][]{Ricci2022, Bicknell1994}. The bolometric luminosity of AGN is $1.4\times10^{43}$ ergs s$^{-1}$ \citep[][]{Gu2007}. Adopting the jet power of $\sim 1.4\times10^{44}$ erg s$^{-1}$ as suggested by \citet{Ricci2022}, which is consistent with measurement using other different methods, radiative power is a factor of $\sim 10$ lower than the bolometric luminosity. Therefore, jet-mode AGN feedback might be the dominant mode of feedback in the galaxy. As mentioned earlier, no significant clumpy UV emission is observed in any other region along the dust filaments, suggesting other mechanisms, such as AGN feedback, in addition to minor mergers, might be playing a role in triggering SF at the UV knot. Simulations suggest winds/jet from the AGN can result in local enhancement in the SF \citep[e.g.,][]{Mukherjee2018, Mondal2021, Mandal2024}. Mechanical feedback from AGN jets has the potential to trigger SF by compressing the gas in the lateral direction as indicated by observations \citep[e.g.,][]{Audibert2023, Mukherjee2018a}. \citet{Audibert2023} also observed the compression and acceleration of gas in the lateral direction due to the interaction of the AGN jet in the Teacup galaxy with the ISM. 

NGC 315 is the brightest galaxy in its group \citep{Chen2012} and is classified as a cool-core galaxy based on X-ray observations \citep{Sun2009}. Elevated SF with SFRs extending to several tens of M$_\odot$ yr$^{-1}$ have been observed in many BCGs \citep[e.g.,][]{Odea2004, Odea2010,Donahue2015,Fogarty2015}. \citet{Odea2004} reported enhanced SFR towards the edge of the radio sources. AGN feedback-driven SF in the cooling gas of BCG and/or condensation of gas uplifted by AGN jet are considered as the mechanisms for triggering SF in these galaxies \citep[e.g.,][]{Donahue2015, Fogarty2015}, which could be a scenario for NGC 315.

The star-forming filaments observed along the jet of Cen A are also associated with dust \citep{Salome2016}. Similar observations are also observed in Cygnus A \citep{Jackson1998}. The origin of gas along the filaments in Cen A and Cygnus A has often been attributed to the external origin, such as the remnant of the merger \citep{Crockett2012, Santoro2015a, Jackson1998}. It is suggested that the SF could have been triggered by shocks after jet propagation in these galaxies \citep{Jackson1998}. This scenario could be the case for NGC 315, where the SF is triggered via AGN feedback in the gas supplied because of minor mergers in the galaxy. As far as $\tau_{\text{dep}}$ is concerned, both merger and AGN induced SF regions can result in lower $\tau_{\text{dep}}$ \citep[e.g.,][]{Gracia2020, Salome2015}.

\subsection{Possible origin of dust}
 
As minor mergers might have been common in NGC 315 \citep{Morganti2009}, dust associated with clouds of minor mergers can cause absorption patches in high-resolution optical observations. The dust in elliptical galaxies has often been associated with merger events \citep[e.g.,][]{Tadhunter2014}.

It is also possible that AGN might be playing a role in the formation/expulsion of dust. The connection between dusty filaments and the central region is reported in earlier studies for other galaxies \citep{Richtler2020, Mathews2013, Temi2007}. \citet{Hirashita2015} and \citet{Hirashita2017} provided a mechanism through which dust can be formed due to AGN feedback. They suggested that injection of the cold gas via minor mergers into the hot gas from AGN feedback can result in the accretion of gas-phase metals, facilitating the growth of dust.

If dust formation is facilitated by the presence of AGN, the coexistence of SF and dust suggests that the dust is long-lived with a sputtering time scale of at least $\sim 10{^7}-10{^8}$ years. The large sputtering time scale could be due to the low density of hot gas and/or the larger grain size of the dust as indicated by equation~(7) of \citet{Hirashita2017}. The low density of the hot gas is indeed suggested by \citet{Chen2012} based on the discrepancy between observed X-ray luminosity and that expected from the velocity dispersion of galaxies in the group of NGC 315. Therefore, our results are also consistent with the dust formation due to the AGN feedback as suggested by \citet{Hirashita2017}. However, further studies of AGN feedback simulation, including dust formation and destruction, are essential, as such studies are currently very limited.

It is also to be noted that the dust filaments are asymmetric in nature, with most of them lying towards the southwest direction. It could also be a case of ram-pressure stripping caused by the motion of NGC 315 through the intragroup medium.

%%%%%%%%%%%%%%%%%%%%%%%%%%%%Summary%%%%%%%%%%%%%%%%%%%%%

\section{Summary}
A distinct knot of UV emission at $\sim 1.7$ kpc towards the southern direction of NGC 315 is detected with UVIT observations. The UVOT colour map supports this UV excess compared to the surrounding regions. We suggest that this patchy and spatially extended UV emission is likely due to ongoing SF in NGC 315. The estimated SFR of $0.23\pm0.10$M$_\odot$ yr$^{-1}$, averaged over 100 Myr, for the UV knot is considerably larger for elliptical galaxies, suggesting that the nuclear region of NGC 315 is actively forming stars. The dust distribution reveals a filamentary structure that extends through the UV SF knot, and the UV knot is located along the filamentary structure extending to the southeast direction. The galaxy does not show signatures of major mergers; therefore, we suggest that minor mergers and/or AGN feedback could be playing a role in triggering SF. Based on this observation, we speculate on possible AGN-triggered SF at the UV knot, where the cool gas could have been supplied via minor mergers, cooling of the gas falling into the central BCG, and/or condensation of the gas uplifted by AGN jet. The formation of dust, possibly facilitated by AGN jets through the accretion of gas-phase metals as suggested by \citet{Hirashita2017}, cannot be ruled out. However, it should be noted that the AGN feedback simulation studies, including the formation and destruction of dust in the radio jet feedback, are very limited. Also, future high-resolution spatially resolved spectroscopic observations are essential to understand the origin of the dust. Further, though we attribute the UV knot emission to SF in NGC 315, additional spectroscopic observations are necessary to confirm if the UV emission is indeed associated with NGC 315.

\section*{Acknowledgements}

We thank the anonymous referee for their valuable comments that helped to improve the quality of the manuscript. The authors thank Dr. Chien Y. Peng for the fruitful discussions regarding \texttt{GALFIT}. BA acknowledges the financial support by DST, Government of India, under the DST-INSPIRE Fellowship (Application Reference Number: DST/INSPIRE/03/2018/000689; INSPIRE Code: IF190146) program. DB thanks the Department of Space, Govt of India, for the financial support under the ISRO Announcement of Opportunity (AO-3) project (grant number No.DS\_2B-13013(2)/2/2021-Sec.2)
The research is based to a significant extent on the results
obtained from the AstroSat mission of the Indian Space Research
Organisation (ISRO), archived at the Indian Space Science Data Centre
(ISSDC). The Payload Operations Centre at IIA processed the UVIT data used here. The UVIT is
built in collaboration between IIA, IUCAA, TIFR, ISRO and CSA. This research is also based on observations made with the {\sl Neil Gehrels Swift Observatory}, obtained from the MAST data archive at the Space Telescope Science Institute, which is operated by the Association of Universities for Research in Astronomy, Inc., under NASA contract NAS 5–26555.
This research is based on observations made with the NASA/ESA {\sl Hubble Space Telescope} obtained from the Space Telescope Science Institute, which is operated by the Association of Universities for Research in Astronomy, Inc., under NASA contract NAS 5–26555. These observations are associated with program 6673. This research is based on observations made with the {\sl Galaxy Evolution Explorer}, obtained from the MAST data archive at the Space Telescope Science Institute, which is operated by the Association of Universities for Research in Astronomy, Inc., under NASA contract NAS 5–26555. This research has made use of data, software and/or web tools obtained from the High Energy Astrophysics Science Archive Research Center (HEASARC), a service of the Astrophysics Science Division at NASA/GSFC and of the Smithsonian Astrophysical Observatory's High Energy Astrophysics Division. This research has made use of the NASA/IPAC Extragalactic Database (NED), which is funded by the National Aeronautics and Space Administration and operated by the California Institute of Technology.
Manipal Centre for Natural Sciences, Centre of Excellence, Manipal Academy of Higher Education (MAHE) is acknowledged for facilities and support. The authors have no conflicts of interest to declare that are relevant to the content of this article.

%%%%%%%%%%%%%%%%%%%%%%%%%%%%%%%%%%%%%%%%%%%%%%%%%%
\section*{Data Availability}
The {\sl AstroSat}-UVIT data used in this work is available at \url{https://astrobrowse.issdc.gov.in/astro_archive/archive/Home.jsp} with observation ID: A11\_101T01\_9000005238. The {\sl Swift}-UVOT data is available at \url{https://heasarc.gsfc.nasa.gov/cgi-bin/W3Browse/swift.pl}. The {\sl GALEX} data is available at \url{https://galex.stsci.edu/gr6/}. The {\sl HST} data used here can be found at \url{http://dx.doi.org/10.17909/m98x-er97}. The remaining data that support the findings of this study are available from the corresponding author upon reasonable request.



\bibliographystyle{mnras}
\bibliography{Astro_ref} % if your bibtex file is called example.bib

@phdthesis{Decleir2019a,
  author       = {{Decleir}, Marjorie},
  language     = {{und}},
  school       = {{Ghent University}},
  title        = {{DustKING - Revealing the dust attenuation in nearby galaxies}},
  year         = {{2019}},
}

@ARTICLE{Decleir2019b,
       author = {{Decleir}, Marjorie and {De Looze}, Ilse and {Boquien}, M{\'e}d{\'e}ric and {Baes}, Maarten and {Verstocken}, Sam and {Calzetti}, Daniela and {Ciesla}, Laure and {Fritz}, Jacopo and {Kennicutt}, Rob and {Nersesian}, Angelos and {Page}, Mathew},
        title = "{Revealing the dust attenuation properties on resolved scales in NGC 628 with SWIFT UVOT data}",
      journal = {\mnras},
         year = 2019,
        month = jun,
       volume = {486},
       number = {1},
        pages = {743-767},
          doi = {10.1093/mnras/stz805},
archivePrefix = {arXiv},
       eprint = {1903.06715},
 primaryClass = {astro-ph.GA},
       adsurl = {https://ui.adsabs.harvard.edu/abs/2019MNRAS.486..743D}
}

@ARTICLE{Postma2017,
       author = {{Postma}, Joseph E. and {Leahy}, Denis},
        title = "{CCDLAB: A Graphical User Interface FITS Image Data Reducer, Viewer, and Canadian UVIT Data Pipeline}",
      journal = {\pasp},
         year = 2017,
        month = nov,
       volume = {129},
       number = {981},
        pages = {115002},
          doi = {10.1088/1538-3873/aa8800},
       adsurl = {https://ui.adsabs.harvard.edu/abs/2017PASP..129k5002P}
}

@ARTICLE{Postma2021,
       author = {{Postma}, Joseph E. and {Leahy}, Denis},
        title = "{UVIT data reduction pipeline: A CCDLAB and UVIT tutorial}",
      journal = {Journal of Astrophysics and Astronomy},
         year = 2021,
        month = oct,
       volume = {42},
       number = {2},
          eid = {30},
        pages = {30},
          doi = {10.1007/s12036-020-09689-w},
       adsurl = {https://ui.adsabs.harvard.edu/abs/2021JApA...42...30P}
}

@ARTICLE{Tandon2017,
   author = {{Tandon}, S.~N. and {Subramaniam}, A. and {Girish}, V. and {Postma}, J. and 
	{Sankarasubramanian}, K. and {Sriram}, S. and {Stalin}, C.~S. and 
	{Mondal}, C. and {Sahu}, S. and {Joseph}, P. and {Hutchings}, J. and 
	{Ghosh}, S.~K. and {Barve}, I.~V. and {George}, K. and {Kamath}, P.~U. and 
	{Kathiravan}, S. and {Kumar}, A. and {Lancelot}, J.~P. and {Leahy}, D. and 
	{Mahesh}, P.~K. and {Mohan}, R. and {Nagabhushana}, S. and {Pati}, A.~K. and 
	{Kameswara Rao}, N. and {Sreedhar}, Y.~H. and {Sreekumar}, P.
	},
    title = "{In-orbit Calibrations of the Ultraviolet Imaging Telescope}",
  journal = {\aj},
archivePrefix = "arXiv",
   eprint = {1705.03715},
 primaryClass = "astro-ph.IM",
     year = 2017,
    month = sep,
   volume = 154,
      eid = {128},
    pages = {128},
      doi = {10.3847/1538-3881/aa8451},
   adsurl = {http://adsabs.harvard.edu/abs/2017AJ....154..128T}
}

@ARTICLE{Roming2005,
       author = {{Roming}, Peter W.~A. and {Kennedy}, Thomas E. and {Mason}, Keith O. and {Nousek}, John A. and {Ahr}, Lindy and {Bingham}, Richard E. and {Broos}, Patrick S. and {Carter}, Mary J. and {Hancock}, Barry K. and {Huckle}, Howard E. and {Hunsberger}, S.~D. and {Kawakami}, Hajime and {Killough}, Ronnie and {Koch}, T. Scott and {McLelland}, Michael K. and {Smith}, Kelly and {Smith}, Philip J. and {Soto}, Juan Carlos and {Boyd}, Patricia T. and {Breeveld}, Alice A. and {Holland}, Stephen T. and {Ivanushkina}, Mariya and {Pryzby}, Michael S. and {Still}, Martin D. and {Stock}, Joseph},
        title = "{The Swift Ultra-Violet/Optical Telescope}",
      journal = {\ssr},
         year = 2005,
        month = oct,
       volume = {120},
       number = {3-4},
        pages = {95-142},
          doi = {10.1007/s11214-005-5095-4},
archivePrefix = {arXiv},
       eprint = {astro-ph/0507413},
 primaryClass = {astro-ph},
       adsurl = {https://ui.adsabs.harvard.edu/abs/2005SSRv..120...95R}
}

@ARTICLE{Peng2002,
       author = {{Peng}, Chien Y. and {Ho}, Luis C. and {Impey}, Chris D. and {Rix}, Hans-Walter},
        title = "{Detailed Structural Decomposition of Galaxy Images}",
      journal = {\aj},
         year = 2002,
        month = jul,
       volume = {124},
       number = {1},
        pages = {266-293},
          doi = {10.1086/340952},
archivePrefix = {arXiv},
       eprint = {astro-ph/0204182},
 primaryClass = {astro-ph},
       adsurl = {https://ui.adsabs.harvard.edu/abs/2002AJ....124..266P}
}

@ARTICLE{Peng2010,
       author = {{Peng}, Chien Y. and {Ho}, Luis C. and {Impey}, Chris D. and {Rix}, Hans-Walter},
        title = "{Detailed Decomposition of Galaxy Images. II. Beyond Axisymmetric Models}",
      journal = {\aj},
         year = 2010,
        month = jun,
       volume = {139},
       number = {6},
        pages = {2097-2129},
          doi = {10.1088/0004-6256/139/6/2097},
archivePrefix = {arXiv},
       eprint = {0912.0731},
 primaryClass = {astro-ph.CO},
       adsurl = {https://ui.adsabs.harvard.edu/abs/2010AJ....139.2097P}
}

@ARTICLE{Kleijn1999,
       author = {{Verdoes Kleijn}, Gijs A. and {Baum}, Stefi A. and {de Zeeuw}, P. Tim and {O'Dea}, Chris P.},
        title = "{Hubble Space Telescope Observations of Nearby Radio-Loud Early-Type Galaxies}",
      journal = {\aj},
         year = 1999,
        month = dec,
       volume = {118},
       number = {6},
        pages = {2592-2617},
          doi = {10.1086/301135},
archivePrefix = {arXiv},
       eprint = {astro-ph/9909256},
 primaryClass = {astro-ph},
       adsurl = {https://ui.adsabs.harvard.edu/abs/1999AJ....118.2592V}
}

@ARTICLE{Capetti2000,
       author = {{Capetti}, A. and {de Ruiter}, H.~R. and {Fanti}, R. and {Morganti}, R. and {Parma}, P. and {Ulrich}, M. -H.},
        title = "{The HST snapshot survey of the B2 sample of low luminosity radio-galaxies: a picture gallery}",
      journal = {\aap},
         year = 2000,
        month = oct,
       volume = {362},
        pages = {871-885},
          doi = {10.48550/arXiv.astro-ph/0009056},
archivePrefix = {arXiv},
       eprint = {astro-ph/0009056},
 primaryClass = {astro-ph},
       adsurl = {https://ui.adsabs.harvard.edu/abs/2000A&A...362..871C}
}

@ARTICLE{Bertin1996,
       author = {{Bertin}, E. and {Arnouts}, S.},
        title = "{SExtractor: Software for source extraction.}",
      journal = {\aaps},
         year = 1996,
        month = jun,
       volume = {117},
        pages = {393-404},
          doi = {10.1051/aas:1996164},
       adsurl = {https://ui.adsabs.harvard.edu/abs/1996A&AS..117..393B}
}

@article{Ananthamoorthy2024,
doi = {10.3847/1538-3881/ad4991},
url = {https://dx.doi.org/10.3847/1538-3881/ad4991},
year = {2024},
month = {jun},
publisher = {The American Astronomical Society},
volume = {168},
number = {1},
pages = {22},
author = {B. Ananthamoorthy and Debbijoy Bhattacharya and P. Sreekumar and Swathi B},
title = {Detection of Faint Sources by the UltraViolet Imaging Telescope Onboard AstroSat Using Poisson Distribution of Background},
journal = {The Astronomical Journal}
}

@article{Ford2013,
doi = {10.1088/0004-637X/770/2/137},
url = {https://dx.doi.org/10.1088/0004-637X/770/2/137},
year = {2013},
month = {jun},
publisher = {The American Astronomical Society},
volume = {770},
number = {2},
pages = {137},
author = {H. Alyson Ford and Joel N. Bregman},
title = {DIRECT DETECTIONS OF YOUNG STARS IN NEARBY ELLIPTICAL GALAXIES*},
journal = {The Astrophysical Journal}
}

@article{Pandey2024,
    author = {Pandey, Divya and Kaviraj, Sugata and Saha, Kanak and Sharma, Saurabh},
    title = "{Star formation exists in all early-type galaxies – evidence from ubiquitous structure in UV images}",
    journal = {Monthly Notices of the Royal Astronomical Society},
    volume = {531},
    number = {2},
    pages = {2223-2236},
    year = {2024},
    month = {05},
    issn = {0035-8711},
    doi = {10.1093/mnras/stae1296},
    url = {https://doi.org/10.1093/mnras/stae1296},
    eprint = {https://academic.oup.com/mnras/article-pdf/531/2/2223/57982675/stae1296.pdf},
}

@article{Finkelman2012,
    author = {Finkelman, Ido and Brosch, Noah and Funes, José G., S. J. and Barway, Sudhanshu and Kniazev, Alexei and Väisänen, Petri},
    title = "{Dust and ionized gas association in E/S0 galaxies with dust lanes: clues to their origin}",
    journal = {Monthly Notices of the Royal Astronomical Society},
    volume = {422},
    number = {2},
    pages = {1384-1393},
    year = {2012},
    month = {04},
    issn = {0035-8711},
    doi = {10.1111/j.1365-2966.2012.20710.x},
    url = {https://doi.org/10.1111/j.1365-2966.2012.20710.x},
    eprint = {https://academic.oup.com/mnras/article-pdf/422/2/1384/3484856/mnras0422-1384.pdf},
}

@ARTICLE{Dokkum1995,
       author = {{van Dokkum}, P.~G. and {Franx}, M.},
        title = "{Dust in the Cores of Early-Type Galaxies}",
      journal = {\aj},
         year = 1995,
        month = nov,
       volume = {110},
        pages = {2027},
          doi = {10.1086/117667},
archivePrefix = {arXiv},
       eprint = {astro-ph/9507101},
 primaryClass = {astro-ph},
       adsurl = {https://ui.adsabs.harvard.edu/abs/1995AJ....110.2027V}
}

@ARTICLE{Richtler2020,
       author = {{Richtler}, T. and {Hilker}, M. and {Iodice}, E.},
        title = "{Dust and gas in the central region of NGC 1316 (Fornax A). Its origin and nature}",
      journal = {\aap},
         year = 2020,
        month = nov,
       volume = {643},
          eid = {A120},
        pages = {A120},
          doi = {10.1051/0004-6361/202038150},
archivePrefix = {arXiv},
       eprint = {2010.01606},
 primaryClass = {astro-ph.GA},
       adsurl = {https://ui.adsabs.harvard.edu/abs/2020A&A...643A.120R}
}

@ARTICLE{Chen2012,
       author = {{Chen}, R. and {Peng}, B. and {Strom}, R.~G. and {Wei}, J.},
        title = "{Galaxy group around giant radio galaxy NGC 315}",
      journal = {\mnras},
         year = 2012,
        month = mar,
       volume = {420},
       number = {3},
        pages = {2715-2726},
          doi = {10.1111/j.1365-2966.2011.20245.x},
       adsurl = {https://ui.adsabs.harvard.edu/abs/2012MNRAS.420.2715C}
}

@ARTICLE{Ricci2022,
       author = {{Ricci}, L. and {Boccardi}, B. and {Nokhrina}, E. and {Perucho}, M. and {MacDonald}, N. and {Mattia}, G. and {Grandi}, P. and {Madika}, E. and {Krichbaum}, T.~P. and {Zensus}, J.~A.},
        title = "{Exploring the disk-jet connection in NGC 315}",
      journal = {\aap},
         year = 2022,
        month = aug,
       volume = {664},
          eid = {A166},
        pages = {A166},
          doi = {10.1051/0004-6361/202243958},
archivePrefix = {arXiv},
       eprint = {2206.12193},
 primaryClass = {astro-ph.HE},
       adsurl = {https://ui.adsabs.harvard.edu/abs/2022A&A...664A.166R}
}

@ARTICLE{Morganti2009,
       author = {{Morganti}, R. and {Peck}, A.~B. and {Oosterloo}, T.~A. and {van Moorsel}, G. and {Capetti}, A. and {Fanti}, R. and {Parma}, P. and {de Ruiter}, H.~R.},
        title = "{Is cold gas fuelling the radio galaxy NGC 315?}",
      journal = {\aap},
         year = 2009,
        month = oct,
       volume = {505},
       number = {2},
        pages = {559-567},
          doi = {10.1051/0004-6361/200912605},
archivePrefix = {arXiv},
       eprint = {0908.3951},
 primaryClass = {astro-ph.CO},
       adsurl = {https://ui.adsabs.harvard.edu/abs/2009A&A...505..559M}
}

@ARTICLE{Trauger1994,
       author = {{Trauger}, John T. and {Ballester}, Gilda E. and {Burrows}, Christopher J. and {Casertano}, Stefano and {Clarke}, John T. and {Crisp}, David and {Evans}, Robin W. and {Gallagher}, John S., III and {Griffiths}, Richard E. and {Hester}, J. Jeff and {Hoessel}, John G. and {Holtzman}, Jon A. and {Krist}, John E. and {Mould}, Jeremy R. and {Scowen}, Paul A. and {Stapelfeldt}, Karl R. and {Watson}, Alan M. and {Westphal}, James A.},
        title = "{The On-Orbit Performance of WFPC2}",
      journal = {\apjl},
         year = 1994,
        month = nov,
       volume = {435},
        pages = {L3},
          doi = {10.1086/187580},
       adsurl = {https://ui.adsabs.harvard.edu/abs/1994ApJ...435L...3T}
}

@ARTICLE{Holtzman1995,
       author = {{Holtzman}, Jon A. and {Hester}, J. Jeff and {Casertano}, Stefano and {Trauger}, John T. and {Watson}, Alan M. and {Ballester}, Gilda E. and {Burrows}, Christopher J. and {Clarke}, John T. and {Crisp}, David and {Evans}, Robin W. and {Gallagher}, John S., III and {Griffiths}, Richard E. and {Hoessel}, John G. and {Matthews}, Lynn D. and {Mould}, Jeremy R. and {Scowen}, Paul A. and {Stapelfeldt}, Karl R. and {Westphal}, James A.},
        title = "{The Performance and Calibration of WFPC2 on the Hubble Space Telescope}",
      journal = {\pasp},
         year = 1995,
        month = feb,
       volume = {107},
        pages = {156},
          doi = {10.1086/133533},
       adsurl = {https://ui.adsabs.harvard.edu/abs/1995PASP..107..156H}
}

@Article{Oppenheimer2021,
AUTHOR = {Oppenheimer, Benjamin D. and Babul, Arif and Bahé, Yannick and Butsky, Iryna S. and McCarthy, Ian G.},
TITLE = {Simulating Groups and the IntraGroup Medium: The Surprisingly Complex and Rich Middle Ground between Clusters and Galaxies},
JOURNAL = {Universe},
VOLUME = {7},
YEAR = {2021},
NUMBER = {7},
ARTICLE-NUMBER = {209},
URL = {https://www.mdpi.com/2218-1997/7/7/209},
ISSN = {2218-1997},
DOI = {10.3390/universe7070209}
}

@ARTICLE{Connel1999,
       author = {{O'Connell}, Robert W.},
        title = "{Far-Ultraviolet Radiation from Elliptical Galaxies}",
      journal = {\araa},
         year = 1999,
        month = jan,
       volume = {37},
        pages = {603-648},
          doi = {10.1146/annurev.astro.37.1.603},
archivePrefix = {arXiv},
       eprint = {astro-ph/9906068},
 primaryClass = {astro-ph},
       adsurl = {https://ui.adsabs.harvard.edu/abs/1999ARA&A..37..603O}
}

@article{Hardcastle2020,
title = {Radio galaxies and feedback from AGN jets},
journal = {New Astronomy Reviews},
volume = {88},
pages = {101539},
year = {2020},
issn = {1387-6473},
doi = {https://doi.org/10.1016/j.newar.2020.101539},
url = {https://www.sciencedirect.com/science/article/pii/S1387647320300166},
author = {M.J. Hardcastle and J.H. Croston}
}

@ARTICLE{Harrison2024,
       author = {{Harrison}, Chris M. and {Ramos Almeida}, Cristina},
        title = "{Observational Tests of Active Galactic Nuclei Feedback: An Overview of Approaches and Interpretation}",
      journal = {Galaxies},
         year = 2024,
        month = apr,
       volume = {12},
       number = {2},
          eid = {17},
        pages = {17},
          doi = {10.3390/galaxies12020017},
archivePrefix = {arXiv},
       eprint = {2404.08050},
 primaryClass = {astro-ph.GA},
       adsurl = {https://ui.adsabs.harvard.edu/abs/2024Galax..12...17H}
}

@ARTICLE{Fabian2012,
       author = {{Fabian}, A.~C.},
        title = "{Observational Evidence of Active Galactic Nuclei Feedback}",
      journal = {\araa},
         year = 2012,
        month = sep,
       volume = {50},
        pages = {455-489},
          doi = {10.1146/annurev-astro-081811-125521},
archivePrefix = {arXiv},
       eprint = {1204.4114},
 primaryClass = {astro-ph.CO},
       adsurl = {https://ui.adsabs.harvard.edu/abs/2012ARA&A..50..455F}
}

@article{Shabala2009,
doi = {10.1088/0004-637X/699/1/525},
url = {https://dx.doi.org/10.1088/0004-637X/699/1/525},
year = {2009},
month = {jun},
publisher = {The American Astronomical Society},
volume = {699},
number = {1},
pages = {525},
author = {Stanislav Shabala and Paul Alexander},
title = {RADIO SOURCE FEEDBACK IN GALAXY EVOLUTION},
journal = {The Astrophysical Journal}
}

@article{Delogu2008,
    author = {Antonuccio-Delogu, V. and Silk, J.},
    title = "{Active galactic nuclei jet-induced feedback in galaxies – I. Suppression of star formation}",
    journal = {Monthly Notices of the Royal Astronomical Society},
    volume = {389},
    number = {4},
    pages = {1750-1762},
    year = {2008},
    month = {09},
    issn = {0035-8711},
    doi = {10.1111/j.1365-2966.2008.13663.x},
    url = {https://doi.org/10.1111/j.1365-2966.2008.13663.x},
    eprint = {https://academic.oup.com/mnras/article-pdf/389/4/1750/3862950/mnras0389-1750.pdf},
}

@article{Mukherjee2018,
    author = {{Mukherjee}, Dipanjan and Bicknell, Geoffrey V and Wagner, Alexander Y and Sutherland, Ralph S and Silk, Joseph},
    title = "{Relativistic jet feedback – III. Feedback on gas discs}",
    journal = {Monthly Notices of the Royal Astronomical Society},
    volume = {479},
    number = {4},
    pages = {5544-5566},
    year = {2018},
    month = {07},
    issn = {0035-8711},
    doi = {10.1093/mnras/sty1776},
    url = {https://doi.org/10.1093/mnras/sty1776},
    eprint = {https://academic.oup.com/mnras/article-pdf/479/4/5544/25218288/sty1776.pdf},
}

@ARTICLE{Boehringer1993,
       author = {{Boehringer}, H. and {Voges}, W. and {Fabian}, A.~C. and {Edge}, A.~C. and {Neumann}, D.~M.},
        title = "{A ROSAT HRI study of the interaction of the X-ray emitting gas and radio lobes of NGC 1275.}",
      journal = {\mnras},
         year = 1993,
        month = oct,
       volume = {264},
        pages = {L25-L28},
          doi = {10.1093/mnras/264.1.L25},
       adsurl = {https://ui.adsabs.harvard.edu/abs/1993MNRAS.264L..25B}
}

@ARTICLE{Dunn2004,
       author = {{Dunn}, R.~J.~H. and {Fabian}, A.~C.},
        title = "{Particle energies and filling fractions of radio bubbles in cluster cores}",
      journal = {\mnras},
         year = 2004,
        month = dec,
       volume = {355},
       number = {3},
        pages = {862-873},
          doi = {10.1111/j.1365-2966.2004.08365.x},
archivePrefix = {arXiv},
       eprint = {astro-ph/0409055},
 primaryClass = {astro-ph},
       adsurl = {https://ui.adsabs.harvard.edu/abs/2004MNRAS.355..862D}
}

@ARTICLE{Salome2016,
       author = {{Salom{\'e}}, Q. and {Salom{\'e}}, P. and {Combes}, F. and {Hamer}, S. and {Heywood}, I.},
        title = "{Star formation efficiency along the radio jet in Centaurus A}",
      journal = {\aap},
         year = 2016,
        month = feb,
       volume = {586},
          eid = {A45},
        pages = {A45},
          doi = {10.1051/0004-6361/201526409},
archivePrefix = {arXiv},
       eprint = {1511.04310},
 primaryClass = {astro-ph.GA},
       adsurl = {https://ui.adsabs.harvard.edu/abs/2016A&A...586A..45S}
}

@ARTICLE{Santoro2015,
       author = {{Santoro}, F. and {Oonk}, J.~B.~R. and {Morganti}, R. and {Oosterloo}, T.~A. and {Tremblay}, G.},
        title = "{The outer filament of Centaurus A as seen by MUSE}",
      journal = {\aap},
         year = 2015,
        month = mar,
       volume = {575},
          eid = {L4},
        pages = {L4},
          doi = {10.1051/0004-6361/201425511},
archivePrefix = {arXiv},
       eprint = {1501.06906},
 primaryClass = {astro-ph.GA},
       adsurl = {https://ui.adsabs.harvard.edu/abs/2015A&A...575L...4S}
}

@ARTICLE{Crockett2012,
       author = {{Crockett}, R. Mark and {Shabala}, Stanislav S. and {Kaviraj}, Sugata and {Antonuccio-Delogu}, Vincenzo and {Silk}, Joseph and {Mutchler}, Max and {O'Connell}, Robert W. and {Rejkuba}, Marina and {Whitmore}, Bradley C. and {Windhorst}, Rogier A.},
        title = "{Triggered star formation in the inner filament of Centaurus A}",
      journal = {\mnras},
         year = 2012,
        month = apr,
       volume = {421},
       number = {2},
        pages = {1603-1623},
          doi = {10.1111/j.1365-2966.2012.20418.x},
archivePrefix = {arXiv},
       eprint = {1201.3369},
 primaryClass = {astro-ph.CO},
       adsurl = {https://ui.adsabs.harvard.edu/abs/2012MNRAS.421.1603C}
}

@ARTICLE{Croft2006,
       author = {{Croft}, Steve and {van Breugel}, Wil and {de Vries}, Wim and {Dopita}, Mike and {Martin}, Chris and {Morganti}, Raffaella and {Neff}, Susan and {Oosterloo}, Tom and {Schiminovich}, David and {Stanford}, S.~A. and {van Gorkom}, Jacqueline},
        title = "{Minkowski's Object: A Starburst Triggered by a Radio Jet, Revisited}",
      journal = {\apj},
         year = 2006,
        month = aug,
       volume = {647},
       number = {2},
        pages = {1040-1055},
          doi = {10.1086/505526},
archivePrefix = {arXiv},
       eprint = {astro-ph/0604557},
 primaryClass = {astro-ph},
       adsurl = {https://ui.adsabs.harvard.edu/abs/2006ApJ...647.1040C}
}

@ARTICLE{Lacy2017,
       author = {{Lacy}, Mark and {Croft}, Steve and {Fragile}, Chris and {Wood}, Sarah and {Nyland}, Kristina},
        title = "{ALMA Observations of the Interaction of a Radio Jet with Molecular Gas in Minkowski's Object}",
      journal = {\apj},
         year = 2017,
        month = apr,
       volume = {838},
       number = {2},
          eid = {146},
        pages = {146},
          doi = {10.3847/1538-4357/aa65d7},
archivePrefix = {arXiv},
       eprint = {1703.03006},
 primaryClass = {astro-ph.GA},
       adsurl = {https://ui.adsabs.harvard.edu/abs/2017ApJ...838..146L}
}

@ARTICLE{Murthy2019,
       author = {{Murthy}, Suma and {Morganti}, Raffaella and {Oosterloo}, Tom and {Schulz}, Robert and {Mukherjee}, Dipanjan and {Wagner}, Alexander Y. and {Bicknell}, Geoffrey and {Prandoni}, Isabella and {Shulevski}, Aleksandar},
        title = "{Feedback from low-luminosity radio galaxies: B2 0258+35}",
      journal = {\aap},
         year = 2019,
        month = sep,
       volume = {629},
          eid = {A58},
        pages = {A58},
          doi = {10.1051/0004-6361/201935931},
archivePrefix = {arXiv},
       eprint = {1908.00374},
 primaryClass = {astro-ph.GA},
       adsurl = {https://ui.adsabs.harvard.edu/abs/2019A&A...629A..58M}
}

@ARTICLE{Lacy1999,
       author = {{Lacy}, Mark and {Ridgway}, Susan E. and {Wold}, Margrethe and {Lilje}, Per B. and {Rawlings}, Steve},
        title = "{Radio-optical alignments in a low radio luminosity sample}",
      journal = {\mnras},
         year = 1999,
        month = aug,
       volume = {307},
       number = {2},
        pages = {420-432},
          doi = {10.1046/j.1365-8711.1999.02640.x},
archivePrefix = {arXiv},
       eprint = {astro-ph/9903314},
 primaryClass = {astro-ph},
       adsurl = {https://ui.adsabs.harvard.edu/abs/1999MNRAS.307..420L}
}

@ARTICLE{Zirm2005,
       author = {{Zirm}, Andrew W. and {Overzier}, R.~A. and {Miley}, G.~K. and {Blakeslee}, J.~P. and {Clampin}, M. and {De Breuck}, C. and {Demarco}, R. and {Ford}, H.~C. and {Hartig}, G.~F. and {Homeier}, N. and {Illingworth}, G.~D. and {Martel}, A.~R. and {R{\"o}ttgering}, H.~J.~A. and {Venemans}, B. and {Ardila}, D.~R. and {Bartko}, F. and {Ben{\'\i}tez}, N. and {Bouwens}, R.~J. and {Bradley}, L.~D. and {Broadhurst}, T.~J. and {Brown}, R.~A. and {Burrows}, C.~J. and {Cheng}, E.~S. and {Cross}, N.~J.~G. and {Feldman}, P.~D. and {Franx}, M. and {Golimowski}, D.~A. and {Goto}, T. and {Gronwall}, C. and {Holden}, B. and {Infante}, L. and {Kimble}, R.~A. and {Krist}, J.~E. and {Lesser}, M.~P. and {Mei}, S. and {Menanteau}, F. and {Meurer}, G.~R. and {Motta}, V. and {Postman}, M. and {Rosati}, P. and {Sirianni}, M. and {Sparks}, W.~B. and {Tran}, H.~D. and {Tsvetanov}, Z.~I. and {White}, R.~L. and {Zheng}, W.},
        title = "{Feedback and Brightest Cluster Galaxy Formation: ACS Observations of the Radio Galaxy TN J1338-1942 at z = 4.1}",
      journal = {\apj},
         year = 2005,
        month = sep,
       volume = {630},
       number = {1},
        pages = {68-81},
          doi = {10.1086/431921},
archivePrefix = {arXiv},
       eprint = {astro-ph/0505610},
 primaryClass = {astro-ph},
       adsurl = {https://ui.adsabs.harvard.edu/abs/2005ApJ...630...68Z}
}

@ARTICLE{Temi2007,
       author = {{Temi}, Pasquale and {Brighenti}, Fabrizio and {Mathews}, William G.},
        title = "{Spitzer Observations of Transient, Extended Dust in Two Elliptical Galaxies: New Evidence of Recent Feedback Energy Release in Galactic Cores}",
      journal = {\apj},
         year = 2007,
        month = sep,
       volume = {666},
       number = {1},
        pages = {222-230},
          doi = {10.1086/520123},
archivePrefix = {arXiv},
       eprint = {0705.3710},
 primaryClass = {astro-ph},
       adsurl = {https://ui.adsabs.harvard.edu/abs/2007ApJ...666..222T}
}

@ARTICLE{Calzetti2000,
       author = {{Calzetti}, Daniela and {Armus}, Lee and {Bohlin}, Ralph C. and {Kinney}, Anne L. and {Koornneef}, Jan and {Storchi-Bergmann}, Thaisa},
        title = "{The Dust Content and Opacity of Actively Star-forming Galaxies}",
      journal = {\apj},
         year = 2000,
        month = apr,
       volume = {533},
       number = {2},
        pages = {682-695},
          doi = {10.1086/308692},
archivePrefix = {arXiv},
       eprint = {astro-ph/9911459},
 primaryClass = {astro-ph},
       adsurl = {https://ui.adsabs.harvard.edu/abs/2000ApJ...533..682C}
}

@ARTICLE{Trager2000,
       author = {{Trager}, S.~C. and {Faber}, S.~M. and {Worthey}, Guy and {Gonz{\'a}lez}, J. Jes{\'u}s},
        title = "{The Stellar Population Histories of Early-Type Galaxies. II. Controlling Parameters of the Stellar Populations}",
      journal = {\aj},
         year = 2000,
        month = jul,
       volume = {120},
       number = {1},
        pages = {165-188},
          doi = {10.1086/301442},
archivePrefix = {arXiv},
       eprint = {astro-ph/0004095},
 primaryClass = {astro-ph},
       adsurl = {https://ui.adsabs.harvard.edu/abs/2000AJ....120..165T}
}

@ARTICLE{Tadhunter2014,
       author = {{Tadhunter}, C. and {Dicken}, D. and {Morganti}, R. and {Konyves}, V. and {Ysard}, N. and {Nesvadba}, N. and {Ramos Almeida}, C.},
        title = "{The dust masses of powerful radio galaxies: clues to the triggering of their activity.}",
      journal = {\mnras},
         year = 2014,
        month = nov,
       volume = {445},
        pages = {L51-L55},
          doi = {10.1093/mnrasl/slu135},
archivePrefix = {arXiv},
       eprint = {1408.3637},
 primaryClass = {astro-ph.GA},
       adsurl = {https://ui.adsabs.harvard.edu/abs/2014MNRAS.445L..51T}
}

@ARTICLE{Pentericci2001,
       author = {{Pentericci}, L. and {McCarthy}, P.~J. and {R{\"o}ttgering}, H.~J.~A. and {Miley}, G.~K. and {van Breugel}, W.~J.~M. and {Fosbury}, R.},
        title = "{NICMOS Observations of High-Redshift Radio Galaxies: Witnessing the Formation of Bright Elliptical Galaxies?}",
      journal = {\apjs},
         year = 2001,
        month = jul,
       volume = {135},
       number = {1},
        pages = {63-85},
          doi = {10.1086/321781},
archivePrefix = {arXiv},
       eprint = {astro-ph/0102323},
 primaryClass = {astro-ph},
       adsurl = {https://ui.adsabs.harvard.edu/abs/2001ApJS..135...63P}
}

@ARTICLE{Ruffa2022,
       author = {{Ruffa}, Ilaria and {Prandoni}, Isabella and {Davis}, Timothy A. and {Laing}, Robert A. and {Paladino}, Rosita and {Casasola}, Viviana and {Parma}, Paola and {Bureau}, Martin},
        title = "{The AGN fuelling/feedback cycle in nearby radio galaxies - IV. Molecular gas conditions and jet-ISM interaction in NGC 3100}",
      journal = {\mnras},
         year = 2022,
        month = mar,
       volume = {510},
       number = {3},
        pages = {4485-4503},
          doi = {10.1093/mnras/stab3541},
archivePrefix = {arXiv},
       eprint = {2112.00755},
 primaryClass = {astro-ph.GA},
       adsurl = {https://ui.adsabs.harvard.edu/abs/2022MNRAS.510.4485R}
}

@ARTICLE{Hirashita2017,
       author = {{Hirashita}, Hiroyuki and {Nozawa}, Takaya},
        title = "{Dust evolution with active galactic nucleus feedback in elliptical galaxies}",
      journal = {\planss},
         year = 2017,
        month = dec,
       volume = {149},
        pages = {45-55},
          doi = {10.1016/j.pss.2017.01.009},
archivePrefix = {arXiv},
       eprint = {1701.07200},
 primaryClass = {astro-ph.GA},
       adsurl = {https://ui.adsabs.harvard.edu/abs/2017P&SS..149...45H}
}

@ARTICLE{Hirashita2015,
       author = {{Hirashita}, Hiroyuki and {Nozawa}, Takaya and {Villaume}, Alexa and {Srinivasan}, Sundar},
        title = "{Dust processing in elliptical galaxies}",
      journal = {\mnras},
         year = 2015,
        month = dec,
       volume = {454},
       number = {2},
        pages = {1620-1633},
          doi = {10.1093/mnras/stv2095},
archivePrefix = {arXiv},
       eprint = {1509.03978},
 primaryClass = {astro-ph.GA},
       adsurl = {https://ui.adsabs.harvard.edu/abs/2015MNRAS.454.1620H}
}

@ARTICLE{Bonfini2014,
       author = {{Bonfini}, Paolo},
        title = "{GALFIT-CORSAIR: Implementing the Core-S{\'e}rsic Model Into GALFIT}",
      journal = {\pasp},
         year = 2014,
        month = oct,
       volume = {126},
       number = {944},
        pages = {935},
          doi = {10.1086/678566},
archivePrefix = {arXiv},
       eprint = {1408.6846},
 primaryClass = {astro-ph.IM},
       adsurl = {https://ui.adsabs.harvard.edu/abs/2014PASP..126..935B}
}

@ARTICLE{Nordon2013,
       author = {{Nordon}, R. and {Lutz}, D. and {Saintonge}, A. and {Berta}, S. and {Wuyts}, S. and {F{\"o}rster Schreiber}, N.~M. and {Genzel}, R. and {Magnelli}, B. and {Poglitsch}, A. and {Popesso}, P. and {Rosario}, D. and {Sturm}, E. and {Tacconi}, L.~J.},
        title = "{The Far-infrared, UV, and Molecular Gas Relation in Galaxies up to z = 2.5}",
      journal = {\apj},
         year = 2013,
        month = jan,
       volume = {762},
       number = {2},
          eid = {125},
        pages = {125},
          doi = {10.1088/0004-637X/762/2/125},
archivePrefix = {arXiv},
       eprint = {1211.5770},
 primaryClass = {astro-ph.CO},
       adsurl = {https://ui.adsabs.harvard.edu/abs/2013ApJ...762..125N}
}

@ARTICLE{Reddy2018,
       author = {{Reddy}, Naveen A. and {Oesch}, Pascal A. and {Bouwens}, Rychard J. and {Montes}, Mireia and {Illingworth}, Garth D. and {Steidel}, Charles C. and {van Dokkum}, Pieter G. and {Atek}, Hakim and {Carollo}, Marcella C. and {Cibinel}, Anna and {Holden}, Brad and {Labb{\'e}}, Ivo and {Magee}, Dan and {Morselli}, Laura and {Nelson}, Erica J. and {Wilkins}, Steve},
        title = "{The HDUV Survey: A Revised Assessment of the Relationship between UV Slope and Dust Attenuation for High-redshift Galaxies}",
      journal = {\apj},
         year = 2018,
        month = jan,
       volume = {853},
       number = {1},
          eid = {56},
        pages = {56},
          doi = {10.3847/1538-4357/aaa3e7},
archivePrefix = {arXiv},
       eprint = {1705.09302},
 primaryClass = {astro-ph.GA},
       adsurl = {https://ui.adsabs.harvard.edu/abs/2018ApJ...853...56R}
}

@ARTICLE{Iglesias2006,
       author = {{Iglesias-P{\'a}ramo}, J. and {Buat}, V. and {Takeuchi}, T.~T. and {Xu}, K. and {Boissier}, S. and {Boselli}, A. and {Burgarella}, D. and {Madore}, B.~F. and {Gil de Paz}, A. and {Bianchi}, L. and {Barlow}, T.~A. and {Byun}, Y. -I. and {Donas}, J. and {Forster}, K. and {Friedman}, P.~G. and {Heckman}, T.~M. and {Jelinski}, P.~N. and {Lee}, Y. -W. and {Malina}, R.~F. and {Martin}, D.~C. and {Milliard}, B. and {Morrissey}, P.~F. and {Neff}, S.~G. and {Rich}, R.~M. and {Schiminovich}, D. and {Seibert}, M. and {Siegmund}, O.~H.~W. and {Small}, T. and {Szalay}, A.~S. and {Welsh}, B.~Y. and {Wyder}, T.~K.},
        title = "{Star Formation in the Nearby Universe: The Ultraviolet and Infrared Points of View}",
      journal = {\apjs},
         year = 2006,
        month = may,
       volume = {164},
       number = {1},
        pages = {38-51},
          doi = {10.1086/502628},
archivePrefix = {arXiv},
       eprint = {astro-ph/0601235},
 primaryClass = {astro-ph},
       adsurl = {https://ui.adsabs.harvard.edu/abs/2006ApJS..164...38I}
}

@ARTICLE{Rubinur2024,
       author = {{Rubinur}, K. and {Das}, M. and {Kharb}, P. and {Yadav}, J. and {Mondal}, C. and {Rahna}, P.~T.},
        title = "{Study of star formation in dual nuclei galaxies using UVIT observations}",
      journal = {\mnras},
         year = 2024,
        month = mar,
       volume = {528},
       number = {3},
        pages = {4432-4450},
          doi = {10.1093/mnras/stae318},
archivePrefix = {arXiv},
       eprint = {2402.02107},
 primaryClass = {astro-ph.GA},
       adsurl = {https://ui.adsabs.harvard.edu/abs/2024MNRAS.528.4432R}
}

@ARTICLE{Plank2016,
       author = {{Planck Collaboration} and {Ade}, P.~A.~R. and {Aghanim}, N. and {Arnaud}, M. and {Ashdown}, M. and {Aumont}, J. and {Baccigalupi}, C. and {Banday}, A.~J. and {Barreiro}, R.~B. and {Bartlett}, J.~G. and {Bartolo}, N. and {Battaner}, E. and {Battye}, R. and {Benabed}, K. and {Beno{\^\i}t}, A. and {Benoit-L{\'e}vy}, A. and {Bernard}, J. -P. and {Bersanelli}, M. and {Bielewicz}, P. and {Bock}, J.~J. and {Bonaldi}, A. and {Bonavera}, L. and {Bond}, J.~R. and {Borrill}, J. and {Bouchet}, F.~R. and {Boulanger}, F. and {Bucher}, M. and {Burigana}, C. and {Butler}, R.~C. and {Calabrese}, E. and {Cardoso}, J. -F. and {Catalano}, A. and {Challinor}, A. and {Chamballu}, A. and {Chary}, R. -R. and {Chiang}, H.~C. and {Chluba}, J. and {Christensen}, P.~R. and {Church}, S. and {Clements}, D.~L. and {Colombi}, S. and {Colombo}, L.~P.~L. and {Combet}, C. and {Coulais}, A. and {Crill}, B.~P. and {Curto}, A. and {Cuttaia}, F. and {Danese}, L. and {Davies}, R.~D. and {Davis}, R.~J. and {de Bernardis}, P. and {de Rosa}, A. and {de Zotti}, G. and {Delabrouille}, J. and {D{\'e}sert}, F. -X. and {Di Valentino}, E. and {Dickinson}, C. and {Diego}, J.~M. and {Dolag}, K. and {Dole}, H. and {Donzelli}, S. and {Dor{\'e}}, O. and {Douspis}, M. and {Ducout}, A. and {Dunkley}, J. and {Dupac}, X. and {Efstathiou}, G. and {Elsner}, F. and {En{\ss}lin}, T.~A. and {Eriksen}, H.~K. and {Farhang}, M. and {Fergusson}, J. and {Finelli}, F. and {Forni}, O. and {Frailis}, M. and {Fraisse}, A.~A. and {Franceschi}, E. and {Frejsel}, A. and {Galeotta}, S. and {Galli}, S. and {Ganga}, K. and {Gauthier}, C. and {Gerbino}, M. and {Ghosh}, T. and {Giard}, M. and {Giraud-H{\'e}raud}, Y. and {Giusarma}, E. and {Gjerl{\o}w}, E. and {Gonz{\'a}lez-Nuevo}, J. and {G{\'o}rski}, K.~M. and {Gratton}, S. and {Gregorio}, A. and {Gruppuso}, A. and {Gudmundsson}, J.~E. and {Hamann}, J. and {Hansen}, F.~K. and {Hanson}, D. and {Harrison}, D.~L. and {Helou}, G. and {Henrot-Versill{\'e}}, S. and {Hern{\'a}ndez-Monteagudo}, C. and {Herranz}, D. and {Hildebrandt}, S.~R. and {Hivon}, E. and {Hobson}, M. and {Holmes}, W.~A. and {Hornstrup}, A. and {Hovest}, W. and {Huang}, Z. and {Huffenberger}, K.~M. and {Hurier}, G. and {Jaffe}, A.~H. and {Jaffe}, T.~R. and {Jones}, W.~C. and {Juvela}, M. and {Keih{\"a}nen}, E. and {Keskitalo}, R. and {Kisner}, T.~S. and {Kneissl}, R. and {Knoche}, J. and {Knox}, L. and {Kunz}, M. and {Kurki-Suonio}, H. and {Lagache}, G. and {L{\"a}hteenm{\"a}ki}, A. and {Lamarre}, J. -M. and {Lasenby}, A. and {Lattanzi}, M. and {Lawrence}, C.~R. and {Leahy}, J.~P. and {Leonardi}, R. and {Lesgourgues}, J. and {Levrier}, F. and {Lewis}, A. and {Liguori}, M. and {Lilje}, P.~B. and {Linden-V{\o}rnle}, M. and {L{\'o}pez-Caniego}, M. and {Lubin}, P.~M. and {Mac{\'\i}as-P{\'e}rez}, J.~F. and {Maggio}, G. and {Maino}, D. and {Mandolesi}, N. and {Mangilli}, A. and {Marchini}, A. and {Maris}, M. and {Martin}, P.~G. and {Martinelli}, M. and {Mart{\'\i}nez-Gonz{\'a}lez}, E. and {Masi}, S. and {Matarrese}, S. and {McGehee}, P. and {Meinhold}, P.~R. and {Melchiorri}, A. and {Melin}, J. -B. and {Mendes}, L. and {Mennella}, A. and {Migliaccio}, M. and {Millea}, M. and {Mitra}, S. and {Miville-Desch{\^e}nes}, M. -A. and {Moneti}, A. and {Montier}, L. and {Morgante}, G. and {Mortlock}, D. and {Moss}, A. and {Munshi}, D. and {Murphy}, J.~A. and {Naselsky}, P. and {Nati}, F. and {Natoli}, P. and {Netterfield}, C.~B. and {N{\o}rgaard-Nielsen}, H.~U. and {Noviello}, F. and {Novikov}, D. and {Novikov}, I. and {Oxborrow}, C.~A. and {Paci}, F. and {Pagano}, L. and {Pajot}, F. and {Paladini}, R. and {Paoletti}, D. and {Partridge}, B. and {Pasian}, F. and {Patanchon}, G. and {Pearson}, T.~J. and {Perdereau}, O. and {Perotto}, L. and {Perrotta}, F. and {Pettorino}, V. and {Piacentini}, F. and {Piat}, M. and {Pierpaoli}, E. and {Pietrobon}, D. and {Plaszczynski}, S. and {Pointecouteau}, E. and {Polenta}, G. and {Popa}, L. and {Pratt}, G.~W. and {Pr{\'e}zeau}, G. and {Prunet}, S. and {Puget}, J. -L. and {Rachen}, J.~P. and {Reach}, W.~T. and {Rebolo}, R. and {Reinecke}, M. and {Remazeilles}, M. and {Renault}, C. and {Renzi}, A. and {Ristorcelli}, I. and {Rocha}, G. and {Rosset}, C. and {Rossetti}, M. and {Roudier}, G. and {Rouill{\'e} d'Orfeuil}, B. and {Rowan-Robinson}, M. and {Rubi{\~n}o-Mart{\'\i}n}, J.~A. and {Rusholme}, B. and {Said}, N. and {Salvatelli}, V. and {Salvati}, L. and {Sandri}, M. and {Santos}, D. and {Savelainen}, M. and {Savini}, G. and {Scott}, D. and {Seiffert}, M.~D. and {Serra}, P. and {Shellard}, E.~P.~S. and {Spencer}, L.~D. and {Spinelli}, M. and {Stolyarov}, V. and {Stompor}, R. and {Sudiwala}, R. and {Sunyaev}, R. and {Sutton}, D. and {Suur-Uski}, A. -S. and {Sygnet}, J. -F. and {Tauber}, J.~A. and {Terenzi}, L. and {Toffolatti}, L. and {Tomasi}, M. and {Tristram}, M. and {Trombetti}, T. and {Tucci}, M. and {Tuovinen}, J. and {T{\"u}rler}, M. and {Umana}, G. and {Valenziano}, L. and {Valiviita}, J. and {Van Tent}, F. and {Vielva}, P. and {Villa}, F. and {Wade}, L.~A. and {Wandelt}, B.~D. and {Wehus}, I.~K. and {White}, M. and {White}, S.~D.~M. and {Wilkinson}, A. and {Yvon}, D. and {Zacchei}, A. and {Zonca}, A.},
        title = "{Planck 2015 results. XIII. Cosmological parameters}",
      journal = {\aap},
         year = 2016,
        month = sep,
       volume = {594},
          eid = {A13},
        pages = {A13},
          doi = {10.1051/0004-6361/201525830},
archivePrefix = {arXiv},
       eprint = {1502.01589},
 primaryClass = {astro-ph.CO},
       adsurl = {https://ui.adsabs.harvard.edu/abs/2016A&A...594A..13P}
}

@ARTICLE{Schlafly2011,
       author = {{Schlafly}, Edward F. and {Finkbeiner}, Douglas P.},
        title = "{Measuring Reddening with Sloan Digital Sky Survey Stellar Spectra and Recalibrating SFD}",
      journal = {\apj},
         year = 2011,
        month = aug,
       volume = {737},
       number = {2},
          eid = {103},
        pages = {103},
          doi = {10.1088/0004-637X/737/2/103},
archivePrefix = {arXiv},
       eprint = {1012.4804},
 primaryClass = {astro-ph.GA},
       adsurl = {https://ui.adsabs.harvard.edu/abs/2011ApJ...737..103S}
}

@ARTICLE{Cardeli1989,
       author = {{Cardelli}, Jason A. and {Clayton}, Geoffrey C. and {Mathis}, John S.},
        title = "{The Relationship between Infrared, Optical, and Ultraviolet Extinction}",
      journal = {\apj},
         year = 1989,
        month = oct,
       volume = {345},
        pages = {245},
          doi = {10.1086/167900},
       adsurl = {https://ui.adsabs.harvard.edu/abs/1989ApJ...345..245C}
}

@ARTICLE{Santoro2015a,
       author = {{Santoro}, F. and {Oonk}, J.~B.~R. and {Morganti}, R. and {Oosterloo}, T.},
        title = "{The jet-ISM interaction in the outer filament of Centaurus A}",
      journal = {\aap},
         year = 2015,
        month = feb,
       volume = {574},
          eid = {A89},
        pages = {A89},
          doi = {10.1051/0004-6361/201425103},
archivePrefix = {arXiv},
       eprint = {1411.4639},
 primaryClass = {astro-ph.GA},
       adsurl = {https://ui.adsabs.harvard.edu/abs/2015A&A...574A..89S}
}

@article{Mathews2013,
doi = {10.1088/0004-637X/768/1/28},
url = {https://dx.doi.org/10.1088/0004-637X/768/1/28},
year = {2013},
month = {apr},
publisher = {The American Astronomical Society},
volume = {768},
number = {1},
pages = {28},
author = {William G. Mathews and Pasquale Temi and Fabrizio Brighenti and Alexandre Amblard},
title = {VARIATIONS OF MID- AND FAR-INFRARED LUMINOSITIES AMONG EARLY-TYPE GALAXIES: RELATION TO STELLAR METALLICITY AND COLD DUST},
journal = {The Astrophysical Journal}
}

@ARTICLE{Jackson1998,
       author = {{Jackson}, Neal and {Tadhunter}, Clive and {Sparks}, William B.},
        title = "{Cygnus A: stars, dust and cones}",
      journal = {\mnras},
         year = 1998,
        month = nov,
       volume = {301},
       number = {1},
        pages = {131-141},
          doi = {10.1046/j.1365-8711.1998.02008.x},
       adsurl = {https://ui.adsabs.harvard.edu/abs/1998MNRAS.301..131J}
}

@ARTICLE{Fragile2017,
       author = {{Fragile}, P. Chris and {Anninos}, Peter and {Croft}, Steve and {Lacy}, Mark and {Witry}, Jason W.~L.},
        title = "{Numerical Simulations of a Jet-Cloud Collision and Starburst: Application to Minkowski{\textquoteright}s Object}",
      journal = {\apj},
         year = 2017,
        month = dec,
       volume = {850},
       number = {2},
          eid = {171},
        pages = {171},
          doi = {10.3847/1538-4357/aa95c6},
archivePrefix = {arXiv},
       eprint = {1701.00024},
 primaryClass = {astro-ph.GA},
       adsurl = {https://ui.adsabs.harvard.edu/abs/2017ApJ...850..171F}
}

@ARTICLE{Fragile2004,
       author = {{Fragile}, P. Chris and {Murray}, Stephen D. and {Anninos}, Peter and {van Breugel}, Wil},
        title = "{Radiative Shock-induced Collapse of Intergalactic Clouds}",
      journal = {\apj},
         year = 2004,
        month = mar,
       volume = {604},
       number = {1},
        pages = {74-87},
          doi = {10.1086/381726},
archivePrefix = {arXiv},
       eprint = {astro-ph/0311298},
 primaryClass = {astro-ph},
       adsurl = {https://ui.adsabs.harvard.edu/abs/2004ApJ...604...74F}
}

@ARTICLE{Wagner2012,
       author = {{Wagner}, A.~Y. and {Bicknell}, G.~V. and {Umemura}, M.},
        title = "{Driving Outflows with Relativistic Jets and the Dependence of Active Galactic Nucleus Feedback Efficiency on Interstellar Medium Inhomogeneity}",
      journal = {\apj},
         year = 2012,
        month = oct,
       volume = {757},
       number = {2},
          eid = {136},
        pages = {136},
          doi = {10.1088/0004-637X/757/2/136},
archivePrefix = {arXiv},
       eprint = {1205.0542},
 primaryClass = {astro-ph.CO},
       adsurl = {https://ui.adsabs.harvard.edu/abs/2012ApJ...757..136W}
}

@ARTICLE{Silk2013,
       author = {{Silk}, Joseph},
        title = "{Unleashing Positive Feedback: Linking the Rates of Star Formation, Supermassive Black Hole Accretion, and Outflows in Distant Galaxies}",
      journal = {\apj},
         year = 2013,
        month = aug,
       volume = {772},
       number = {2},
          eid = {112},
        pages = {112},
          doi = {10.1088/0004-637X/772/2/112},
archivePrefix = {arXiv},
       eprint = {1305.5840},
 primaryClass = {astro-ph.CO},
       adsurl = {https://ui.adsabs.harvard.edu/abs/2013ApJ...772..112S}
}

@ARTICLE{Bridle1976,
       author = {{Bridle}, A.~H. and {Davis}, M.~M. and {Meloy}, D.~A. and {Fomalont}, E.~B. and {Strom}, R.~G. and {Willis}, A.~G.},
        title = "{Giant radio galaxy NGC 315.}",
      journal = {\nat},
         year = 1976,
        month = jul,
       volume = {262},
        pages = {179-182},
          doi = {10.1038/262179a0},
       adsurl = {https://ui.adsabs.harvard.edu/abs/1976Natur.262..179B}
}

@ARTICLE{Fanaroff1974,
       author = {{Fanaroff}, B.~L. and {Riley}, J.~M.},
        title = "{The morphology of extragalactic radio sources of high and low luminosity}",
      journal = {\mnras},
         year = 1974,
        month = may,
       volume = {167},
        pages = {31P-36P},
          doi = {10.1093/mnras/167.1.31P},
       adsurl = {https://ui.adsabs.harvard.edu/abs/1974MNRAS.167P..31F}
}

@ARTICLE{Mukherjee2018a,
       author = {{Mukherjee}, Dipanjan and {Wagner}, Alexander Y. and {Bicknell}, Geoffrey V. and {Morganti}, Raffaella and {Oosterloo}, Tom and {Nesvadba}, Nicole and {Sutherland}, Ralph S.},
        title = "{The jet-ISM interactions in IC 5063}",
      journal = {\mnras},
         year = 2018,
        month = may,
       volume = {476},
       number = {1},
        pages = {80-95},
          doi = {10.1093/mnras/sty067},
archivePrefix = {arXiv},
       eprint = {1801.06875},
 primaryClass = {astro-ph.HE},
       adsurl = {https://ui.adsabs.harvard.edu/abs/2018MNRAS.476...80M}
}

@ARTICLE{Tomar2021,
       author = {{Tomar}, Gunjan and {Gupta}, Nayantara and {Prince}, Raj},
        title = "{Broadband Modeling of Low-luminosity Active Galactic Nuclei Detected in Gamma Rays}",
      journal = {\apj},
         year = 2021,
        month = oct,
       volume = {919},
       number = {2},
          eid = {137},
        pages = {137},
          doi = {10.3847/1538-4357/ac1588},
archivePrefix = {arXiv},
       eprint = {2107.08256},
 primaryClass = {astro-ph.HE},
       adsurl = {https://ui.adsabs.harvard.edu/abs/2021ApJ...919..137T}
}

@ARTICLE{Boizelle2021,
       author = {{Boizelle}, Benjamin D. and {Walsh}, Jonelle L. and {Barth}, Aaron J. and {Buote}, David A. and {Baker}, Andrew J. and {Darling}, Jeremy and {Ho}, Luis C. and {Cohn}, Jonathan and {Kabasares}, Kyle M.},
        title = "{Black Hole Mass Measurements of Radio Galaxies NGC 315 and NGC 4261 Using ALMA CO Observations}",
      journal = {\apj},
         year = 2021,
        month = feb,
       volume = {908},
       number = {1},
          eid = {19},
        pages = {19},
          doi = {10.3847/1538-4357/abd24d},
archivePrefix = {arXiv},
       eprint = {2012.04669},
 primaryClass = {astro-ph.GA},
       adsurl = {https://ui.adsabs.harvard.edu/abs/2021ApJ...908...19B}
}

@ARTICLE{Goudfrooij1994,
       author = {{Goudfrooij}, P. and {Hansen}, L. and {Jorgensen}, H.~E. and {Norgaard-Nielsen}, H.~U.},
        title = "{Interstellar matter in Shapley-Ames elliptical galaxies. II. The distribution of dust and ionized gas}",
      journal = {\aaps},
         year = 1994,
        month = jun,
       volume = {105},
        pages = {341-383},
       adsurl = {https://ui.adsabs.harvard.edu/abs/1994A&AS..105..341G}
}

@ARTICLE{Kroupa2001,
       author = {{Kroupa}, Pavel},
        title = "{On the variation of the initial mass function}",
      journal = {\mnras},
         year = 2001,
        month = apr,
       volume = {322},
       number = {2},
        pages = {231-246},
          doi = {10.1046/j.1365-8711.2001.04022.x},
archivePrefix = {arXiv},
       eprint = {astro-ph/0009005},
 primaryClass = {astro-ph},
       adsurl = {https://ui.adsabs.harvard.edu/abs/2001MNRAS.322..231K}
}

@ARTICLE{Salpeter1955,
       author = {{Salpeter}, Edwin E.},
        title = "{The Luminosity Function and Stellar Evolution.}",
      journal = {\apj},
         year = 1955,
        month = jan,
       volume = {121},
        pages = {161},
          doi = {10.1086/145971},
       adsurl = {https://ui.adsabs.harvard.edu/abs/1955ApJ...121..161S}
}

@ARTICLE{Audibert2023,
       author = {{Audibert}, A. and {Ramos Almeida}, C. and {Garc{\'\i}a-Burillo}, S. and {Combes}, F. and {Bischetti}, M. and {Meenakshi}, M. and {Mukherjee}, D. and {Bicknell}, G. and {Wagner}, A.~Y.},
        title = "{Jet-induced molecular gas excitation and turbulence in the Teacup}",
      journal = {\aap},
         year = 2023,
        month = mar,
       volume = {671},
          eid = {L12},
        pages = {L12},
          doi = {10.1051/0004-6361/202345964},
archivePrefix = {arXiv},
       eprint = {2302.13884},
 primaryClass = {astro-ph.GA},
       adsurl = {https://ui.adsabs.harvard.edu/abs/2023A&A...671L..12A}
}

@ARTICLE{Masegosa2011,
       author = {{Masegosa}, J. and {M{\'a}rquez}, I. and {Ramirez}, A. and {Gonz{\'a}lez-Mart{\'\i}n}, O.},
        title = "{The nature of nuclear H$_{{\ensuremath{\alpha}}}$ emission in LINERs}",
      journal = {\aap},
         year = 2011,
        month = mar,
       volume = {527},
          eid = {A23},
        pages = {A23},
          doi = {10.1051/0004-6361/201015047},
archivePrefix = {arXiv},
       eprint = {1011.0865},
 primaryClass = {astro-ph.CO},
       adsurl = {https://ui.adsabs.harvard.edu/abs/2011A&A...527A..23M}
}

@ARTICLE{Croton2006,
       author = {{Croton}, Darren J. and {Springel}, Volker and {White}, Simon D.~M. and {De Lucia}, G. and {Frenk}, C.~S. and {Gao}, L. and {Jenkins}, A. and {Kauffmann}, G. and {Navarro}, J.~F. and {Yoshida}, N.},
        title = "{The many lives of active galactic nuclei: cooling flows, black holes and the luminosities and colours of galaxies}",
      journal = {\mnras},
         year = 2006,
        month = jan,
       volume = {365},
       number = {1},
        pages = {11-28},
          doi = {10.1111/j.1365-2966.2005.09675.x},
archivePrefix = {arXiv},
       eprint = {astro-ph/0508046},
 primaryClass = {astro-ph},
       adsurl = {https://ui.adsabs.harvard.edu/abs/2006MNRAS.365...11C}
}

@BOOK{RC3,
       author = {{de Vaucouleurs}, Gerard and {de Vaucouleurs}, Antoinette and {Corwin}, Herold G., Jr. and {Buta}, Ronald J. and {Paturel}, Georges and {Fouque}, Pascal},
        title = "{Third Reference Catalogue of Bright Galaxies}",
         year = 1991,
       adsurl = {https://ui.adsabs.harvard.edu/abs/1991rc3..book.....D}
}

@ARTICLE{Tandon2020,
       author = {{Tandon}, S.~N. and {Postma}, J. and {Joseph}, P. and {Devaraj}, A. and {Subramaniam}, A. and {Barve}, I.~V. and {George}, K. and {Ghosh}, S.~K. and {Girish}, V. and {Hutchings}, J.~B. and {Kamath}, P.~U. and {Kathiravan}, S. and {Kumar}, A. and {Lancelot}, J.~P. and {Leahy}, D. and {Mahesh}, P.~K. and {Mohan}, R. and {Nagabhushana}, S. and {Pati}, A.~K. and {Rao}, N. Kameswara and {Sankarasubramanian}, K. and {Sriram}, S. and {Stalin}, C.~S.},
        title = "{Additional Calibration of the Ultraviolet Imaging Telescope on Board AstroSat}",
      journal = {\aj},
         year = 2020,
        month = apr,
       volume = {159},
       number = {4},
          eid = {158},
        pages = {158},
          doi = {10.3847/1538-3881/ab72a3},
archivePrefix = {arXiv},
       eprint = {2002.01159},
 primaryClass = {astro-ph.IM},
       adsurl = {https://ui.adsabs.harvard.edu/abs/2020AJ....159..158T}
}

@ARTICLE{Morrissey2007,
       author = {{Morrissey}, Patrick and {Conrow}, Tim and {Barlow}, Tom A. and {Small}, Todd and {Seibert}, Mark and {Wyder}, Ted K. and {Budav{\'a}ri}, Tam{\'a}s and {Arnouts}, Stephane and {Friedman}, Peter G. and {Forster}, Karl and {Martin}, D. Christopher and {Neff}, Susan G. and {Schiminovich}, David and {Bianchi}, Luciana and {Donas}, Jos{\'e} and {Heckman}, Timothy M. and {Lee}, Young-Wook and {Madore}, Barry F. and {Milliard}, Bruno and {Rich}, R. Michael and {Szalay}, Alex S. and {Welsh}, Barry Y. and {Yi}, Sukyoung K.},
        title = "{The Calibration and Data Products of GALEX}",
      journal = {\apjs},
         year = 2007,
        month = dec,
       volume = {173},
       number = {2},
        pages = {682-697},
          doi = {10.1086/520512},
archivePrefix = {arXiv},
       eprint = {0706.0755},
 primaryClass = {astro-ph},
       adsurl = {https://ui.adsabs.harvard.edu/abs/2007ApJS..173..682M}
}

@ARTICLE{Leitherer1999,
       author = {{Leitherer}, Claus and {Schaerer}, Daniel and {Goldader}, Jeffrey D. and {Delgado}, Rosa M. Gonz{\'a}lez and {Robert}, Carmelle and {Kune}, Denis Foo and {de Mello}, Du{\'\i}lia F. and {Devost}, Daniel and {Heckman}, Timothy M.},
        title = "{Starburst99: Synthesis Models for Galaxies with Active Star Formation}",
      journal = {\apjs},
         year = 1999,
        month = jul,
       volume = {123},
       number = {1},
        pages = {3-40},
          doi = {10.1086/313233},
archivePrefix = {arXiv},
       eprint = {astro-ph/9902334},
 primaryClass = {astro-ph},
       adsurl = {https://ui.adsabs.harvard.edu/abs/1999ApJS..123....3L}
}

@ARTICLE{Martin2005,
       author = {{Martin}, D. Christopher and {Fanson}, James and {Schiminovich}, David and {Morrissey}, Patrick and {Friedman}, Peter G. and {Barlow}, Tom A. and {Conrow}, Tim and {Grange}, Robert and {Jelinsky}, Patrick N. and {Milliard}, Bruno and {Siegmund}, Oswald H.~W. and {Bianchi}, Luciana and {Byun}, Yong-Ik and {Donas}, Jose and {Forster}, Karl and {Heckman}, Timothy M. and {Lee}, Young-Wook and {Madore}, Barry F. and {Malina}, Roger F. and {Neff}, Susan G. and {Rich}, R. Michael and {Small}, Todd and {Surber}, Frank and {Szalay}, Alex S. and {Welsh}, Barry and {Wyder}, Ted K.},
        title = "{The Galaxy Evolution Explorer: A Space Ultraviolet Survey Mission}",
      journal = {\apjl},
         year = 2005,
        month = jan,
       volume = {619},
       number = {1},
        pages = {L1-L6},
          doi = {10.1086/426387},
archivePrefix = {arXiv},
       eprint = {astro-ph/0411302},
 primaryClass = {astro-ph},
       adsurl = {https://ui.adsabs.harvard.edu/abs/2005ApJ...619L...1M}
}

@ARTICLE{Worrall2007,
       author = {{Worrall}, D.~M. and {Birkinshaw}, M. and {Laing}, R.~A. and {Cotton}, W.~D. and {Bridle}, A.~H.},
        title = "{The inner jet of radio galaxy NGC 315 as observed with Chandra and the Very Large Array}",
      journal = {\mnras},
         year = 2007,
        month = sep,
       volume = {380},
       number = {1},
        pages = {2-14},
          doi = {10.1111/j.1365-2966.2007.11998.x},
archivePrefix = {arXiv},
       eprint = {0705.4100},
 primaryClass = {astro-ph},
       adsurl = {https://ui.adsabs.harvard.edu/abs/2007MNRAS.380....2W}
}

@ARTICLE{Zhang2008,
       author = {{Zhang}, Y. and {Gu}, Q. -S. and {Ho}, L.~C.},
        title = "{Stellar and dust properties of local elliptical galaxies: clues to the onset of nuclear activity}",
      journal = {\aap},
         year = 2008,
        month = aug,
       volume = {487},
       number = {1},
        pages = {177-183},
          doi = {10.1051/0004-6361:200809660},
archivePrefix = {arXiv},
       eprint = {0806.2189},
 primaryClass = {astro-ph},
       adsurl = {https://ui.adsabs.harvard.edu/abs/2008A&A...487..177Z}
}

@ARTICLE{Vazquez2005,
       author = {{V{\'a}zquez}, Gerardo A. and {Leitherer}, Claus},
        title = "{Optimization of Starburst99 for Intermediate-Age and Old Stellar Populations}",
      journal = {\apj},
         year = 2005,
        month = mar,
       volume = {621},
       number = {2},
        pages = {695-717},
          doi = {10.1086/427866},
archivePrefix = {arXiv},
       eprint = {astro-ph/0412491},
 primaryClass = {astro-ph},
       adsurl = {https://ui.adsabs.harvard.edu/abs/2005ApJ...621..695V}
}

@ARTICLE{Kolokythas2022,
       author = {{Kolokythas}, Konstantinos and {Vaddi}, Sravani and {O'Sullivan}, Ewan and {Loubser}, Ilani and {Babul}, Arif and {Raychaudhury}, Somak and {Lagos}, Patricio and {Jarrett}, Thomas H.},
        title = "{The Complete Local-Volume Groups Sample - IV. Star formation and gas content in group-dominant galaxies}",
      journal = {\mnras},
         year = 2022,
        month = mar,
       volume = {510},
       number = {3},
        pages = {4191-4207},
          doi = {10.1093/mnras/stab3699},
archivePrefix = {arXiv},
       eprint = {2112.08498},
 primaryClass = {astro-ph.GA},
       adsurl = {https://ui.adsabs.harvard.edu/abs/2022MNRAS.510.4191K}
}

@ARTICLE{Kaviraj2007,
       author = {{Kaviraj}, S. and {Schawinski}, K. and {Devriendt}, J.~E.~G. and {Ferreras}, I. and {Khochfar}, S. and {Yoon}, S. -J. and {Yi}, S.~K. and {Deharveng}, J. -M. and {Boselli}, A. and {Barlow}, T. and {Conrow}, T. and {Forster}, K. and {Friedman}, P.~G. and {Martin}, D.~C. and {Morrissey}, P. and {Neff}, S. and {Schiminovich}, D. and {Seibert}, M. and {Small}, T. and {Wyder}, T. and {Bianchi}, L. and {Donas}, J. and {Heckman}, T. and {Lee}, Y. -W. and {Madore}, B. and {Milliard}, B. and {Rich}, R.~M. and {Szalay}, A.},
        title = "{UV-Optical Colors As Probes of Early-Type Galaxy Evolution}",
      journal = {\apjs},
         year = 2007,
        month = dec,
       volume = {173},
       number = {2},
        pages = {619-642},
          doi = {10.1086/516633},
archivePrefix = {arXiv},
       eprint = {astro-ph/0601029},
 primaryClass = {astro-ph},
       adsurl = {https://ui.adsabs.harvard.edu/abs/2007ApJS..173..619K}
}

@INPROCEEDINGS{Yi2008,
       author = {{Yi}, S.~K.},
        title = "{The Current Understanding on the UV Upturn}",
    booktitle = {Hot Subdwarf Stars and Related Objects},
         year = 2008,
       editor = {{Heber}, U. and {Jeffery}, C.~S. and {Napiwotzki}, R.},
       series = {Astronomical Society of the Pacific Conference Series},
       volume = {392},
        month = jan,
        pages = {3},
          doi = {10.48550/arXiv.0808.0254},
archivePrefix = {arXiv},
       eprint = {0808.0254},
 primaryClass = {astro-ph},
       adsurl = {https://ui.adsabs.harvard.edu/abs/2008ASPC..392....3Y}
}

@ARTICLE{Mondal2021,
       author = {{Mandal}, Ankush and {Mukherjee}, Dipanjan and {Federrath}, Christoph and {Nesvadba}, Nicole P.~H. and {Bicknell}, Geoffrey V. and {Wagner}, Alexander Y. and {Meenakshi}, Moun},
        title = "{Impact of relativistic jets on the star formation rate: a turbulence-regulated framework}",
      journal = {\mnras},
         year = 2021,
        month = dec,
       volume = {508},
       number = {4},
        pages = {4738-4757},
          doi = {10.1093/mnras/stab2822},
archivePrefix = {arXiv},
       eprint = {2109.13654},
 primaryClass = {astro-ph.GA},
       adsurl = {https://ui.adsabs.harvard.edu/abs/2021MNRAS.508.4738M}
}

@ARTICLE{Leroy2013,
       author = {{Leroy}, Adam K. and {Walter}, Fabian and {Sandstrom}, Karin and {Schruba}, Andreas and {Munoz-Mateos}, Juan-Carlos and {Bigiel}, Frank and {Bolatto}, Alberto and {Brinks}, Elias and {de Blok}, W.~J.~G. and {Meidt}, Sharon and {Rix}, Hans-Walter and {Rosolowsky}, Erik and {Schinnerer}, Eva and {Schuster}, Karl-Friedrich and {Usero}, Antonio},
        title = "{Molecular Gas and Star Formation in nearby Disk Galaxies}",
      journal = {\aj},
         year = 2013,
        month = aug,
       volume = {146},
       number = {2},
          eid = {19},
        pages = {19},
          doi = {10.1088/0004-6256/146/2/19},
archivePrefix = {arXiv},
       eprint = {1301.2328},
 primaryClass = {astro-ph.CO},
       adsurl = {https://ui.adsabs.harvard.edu/abs/2013AJ....146...19L}
}

@ARTICLE{Gu2007,
       author = {{Gu}, Q. -S. and {Huang}, J. -S. and {Wilson}, G. and {Fazio}, G.~G.},
        title = "{Direct Evidence from Spitzer for a Low-Luminosity AGN at the Center of the Elliptical Galaxy NGC 315}",
      journal = {\apjl},
         year = 2007,
        month = dec,
       volume = {671},
       number = {2},
        pages = {L105-L108},
          doi = {10.1086/525018},
archivePrefix = {arXiv},
       eprint = {0711.0051},
 primaryClass = {astro-ph},
       adsurl = {https://ui.adsabs.harvard.edu/abs/2007ApJ...671L.105G}
}

@book{Krumholz2017,
author = {Krumholz, Mark R},
title = {Star Formation},
publisher = {WORLD SCIENTIFIC},
year = {2017},
doi = {10.1142/10091},
address = {},
edition   = {}
}

@ARTICLE{Gracia2020,
       author = {{D{\'\i}az-Garc{\'\i}a}, S. and {Knapen}, J.~H.},
        title = "{Gas fractions and depletion times in galaxies with different degrees of interaction}",
      journal = {\aap},
         year = 2020,
        month = mar,
       volume = {635},
          eid = {A197},
        pages = {A197},
          doi = {10.1051/0004-6361/201937384},
archivePrefix = {arXiv},
       eprint = {2002.09257},
 primaryClass = {astro-ph.GA},
       adsurl = {https://ui.adsabs.harvard.edu/abs/2020A&A...635A.197D}
}

@ARTICLE{Salome2015,
       author = {{Salom{\'e}}, Q. and {Salom{\'e}}, P. and {Combes}, F.},
        title = "{Jet-induced star formation in 3C 285 and Minkowski's Object}",
      journal = {\aap},
         year = 2015,
        month = feb,
       volume = {574},
          eid = {A34},
        pages = {A34},
          doi = {10.1051/0004-6361/201424932},
archivePrefix = {arXiv},
       eprint = {1410.8367},
 primaryClass = {astro-ph.GA},
       adsurl = {https://ui.adsabs.harvard.edu/abs/2015A&A...574A..34S}
}

@ARTICLE{Narayanan2014,
       author = {{Narayanan}, Desika and {Krumholz}, Mark R.},
        title = "{A theory for the excitation of CO in star-forming galaxies}",
      journal = {\mnras},
         year = 2014,
        month = aug,
       volume = {442},
       number = {2},
        pages = {1411-1428},
          doi = {10.1093/mnras/stu834},
archivePrefix = {arXiv},
       eprint = {1401.2998},
 primaryClass = {astro-ph.GA},
       adsurl = {https://ui.adsabs.harvard.edu/abs/2014MNRAS.442.1411N}
}

@ARTICLE{Davis2015,
       author = {{Davis}, Timothy A. and {Rowlands}, Kate and {Allison}, James R. and {Shabala}, Stanislav S. and {Ting}, Yuan-Sen and {Lagos}, Claudia del P. and {Kaviraj}, Sugata and {Bourne}, Nathan and {Dunne}, Loretta and {Eales}, Steve and {Ivison}, Rob. J. and {Maddox}, Steve and {Smith}, Daniel J.~B. and {Smith}, Matthew W.~L. and {Temi}, Pasquale},
        title = "{Molecular and atomic gas in dust lane early-type galaxies - I. Low star formation efficiencies in minor merger remnants}",
      journal = {\mnras},
         year = 2015,
        month = jun,
       volume = {449},
       number = {4},
        pages = {3503-3516},
          doi = {10.1093/mnras/stv597},
archivePrefix = {arXiv},
       eprint = {1503.05162},
 primaryClass = {astro-ph.GA},
       adsurl = {https://ui.adsabs.harvard.edu/abs/2015MNRAS.449.3503D}
}

@ARTICLE{Kaviraj2011,
       author = {{Kaviraj}, Sugata and {Tan}, Kok-Meng and {Ellis}, Richard S. and {Silk}, Joseph},
        title = "{A coincidence of disturbed morphology and blue UV colour: minor-merger-driven star formation in early-type galaxies at z{\ensuremath{\sim}} 0.6}",
      journal = {\mnras},
         year = 2011,
        month = mar,
       volume = {411},
       number = {4},
        pages = {2148-2160},
          doi = {10.1111/j.1365-2966.2010.17754.x},
archivePrefix = {arXiv},
       eprint = {1001.2141},
 primaryClass = {astro-ph.CO},
       adsurl = {https://ui.adsabs.harvard.edu/abs/2011MNRAS.411.2148K}
}

@ARTICLE{Yi2005,
       author = {{Yi}, S.~K. and {Yoon}, S. -J. and {Kaviraj}, S. and {Deharveng}, J. -M. and {Rich}, R.~M. and {Salim}, S. and {Boselli}, A. and {Lee}, Y. -W. and {Ree}, C.~H. and {Sohn}, Y. -J. and {Rey}, S. -C. and {Lee}, J. -W. and {Rhee}, J. and {Bianchi}, L. and {Byun}, Y. -I. and {Donas}, J. and {Friedman}, P.~G. and {Heckman}, T.~M. and {Jelinsky}, P. and {Madore}, B.~F. and {Malina}, R. and {Martin}, D.~C. and {Milliard}, B. and {Morrissey}, P. and {Neff}, S. and {Schiminovich}, D. and {Siegmund}, O. and {Small}, T. and {Szalay}, A.~S. and {Jee}, M.~J. and {Kim}, S. -W. and {Barlow}, T. and {Forster}, K. and {Welsh}, B. and {Wyder}, T.~K.},
        title = "{Galaxy Evolution Explorer Ultraviolet Color-Magnitude Relations and Evidence of Recent Star Formation in Early-Type Galaxies}",
      journal = {\apjl},
         year = 2005,
        month = jan,
       volume = {619},
       number = {1},
        pages = {L111-L114},
          doi = {10.1086/422811},
archivePrefix = {arXiv},
       eprint = {astro-ph/0411327},
 primaryClass = {astro-ph},
       adsurl = {https://ui.adsabs.harvard.edu/abs/2005ApJ...619L.111Y}
}

@ARTICLE{Hani2020,
       author = {{Hani}, Maan H. and {Gosain}, Hayman and {Ellison}, Sara L. and {Patton}, David R. and {Torrey}, Paul},
        title = "{Interacting galaxies in the IllustrisTNG simulations - II: star formation in the post-merger stage}",
      journal = {\mnras},
         year = 2020,
        month = apr,
       volume = {493},
       number = {3},
        pages = {3716-3731},
          doi = {10.1093/mnras/staa459},
archivePrefix = {arXiv},
       eprint = {2001.04472},
 primaryClass = {astro-ph.GA},
       adsurl = {https://ui.adsabs.harvard.edu/abs/2020MNRAS.493.3716H}
}

@ARTICLE{Matteo2007,
       author = {{Di Matteo}, P. and {Combes}, F. and {Melchior}, A. -L. and {Semelin}, B.},
        title = "{Star formation efficiency in galaxy interactions and mergers: a statistical study}",
      journal = {\aap},
         year = 2007,
        month = jun,
       volume = {468},
       number = {1},
        pages = {61-81},
          doi = {10.1051/0004-6361:20066959},
archivePrefix = {arXiv},
       eprint = {astro-ph/0703212},
 primaryClass = {astro-ph},
       adsurl = {https://ui.adsabs.harvard.edu/abs/2007A&A...468...61D}
}

@ARTICLE{Mathis1990,
       author = {{Mathis}, John S.},
        title = "{Interstellar dust and extinction.}",
      journal = {\araa},
         year = 1990,
        month = jan,
       volume = {28},
        pages = {37-70},
          doi = {10.1146/annurev.aa.28.090190.000345},
       adsurl = {https://ui.adsabs.harvard.edu/abs/1990ARA&A..28...37M}
}

@ARTICLE{Brown1989,
       author = {{Brown}, L.~M.~J. and {Robson}, E.~I. and {Gear}, W.~K. and {Smith}, M.~G.},
        title = "{Multifrequency Observations of Blazars. IV. The Variability of the Radio to Ultraviolet Continuum}",
      journal = {\apj},
         year = 1989,
        month = may,
       volume = {340},
        pages = {150},
          doi = {10.1086/167381},
       adsurl = {https://ui.adsabs.harvard.edu/abs/1989ApJ...340..150B}
}

@ARTICLE{Diego2010,
       author = {{de Diego}, Jos{\'e} A.},
        title = "{Testing Tests on Active Galactic Nucleus Microvariability}",
      journal = {\aj},
         year = 2010,
        month = mar,
       volume = {139},
       number = {3},
        pages = {1269-1282},
          doi = {10.1088/0004-6256/139/3/1269},
archivePrefix = {arXiv},
       eprint = {1001.2543},
 primaryClass = {astro-ph.CO},
       adsurl = {https://ui.adsabs.harvard.edu/abs/2010AJ....139.1269D}
}

@ARTICLE{Nandra1997,
       author = {{Nandra}, K. and {George}, I.~M. and {Mushotzky}, R.~F. and {Turner}, T.~J. and {Yaqoob}, T.},
        title = "{ASCA Observations of Seyfert 1 Galaxies. I. Data Analysis, Imaging, and Timing}",
      journal = {\apj},
         year = 1997,
        month = feb,
       volume = {476},
       number = {1},
        pages = {70-82},
          doi = {10.1086/303600},
       adsurl = {https://ui.adsabs.harvard.edu/abs/1997ApJ...476...70N}
}

@ARTICLE{Rose2007,
       author = {{Rose}, Michael B. and {Hintz}, Eric G.},
        title = "{A Search for Low-Amplitude Variability in Six Open Clusters Using the Robust Median Statistic}",
      journal = {\aj},
         year = 2007,
        month = nov,
       volume = {134},
       number = {5},
        pages = {2067-2078},
          doi = {10.1086/522963},
       adsurl = {https://ui.adsabs.harvard.edu/abs/2007AJ....134.2067R}
}

@ARTICLE{Joseph2022,
       author = {{Joseph}, Prajwel and {Sreekumar}, P. and {Stalin}, C.~S. and {Paul}, K.~T. and {Mondal}, Chayan and {George}, Koshy and {Mathew}, Blesson},
        title = "{UVIT view of Centaurus A: a detailed study on positive AGN feedback}",
      journal = {\mnras},
         year = 2022,
        month = oct,
       volume = {516},
       number = {2},
        pages = {2300-2313},
          doi = {10.1093/mnras/stac2388},
archivePrefix = {arXiv},
       eprint = {2208.10209},
 primaryClass = {astro-ph.GA},
       adsurl = {https://ui.adsabs.harvard.edu/abs/2022MNRAS.516.2300J}
}

@ARTICLE{Kennicutt1998,
       author = {{Kennicutt}, Jr., Robert C.},
        title = "{Star Formation in Galaxies Along the Hubble Sequence}",
      journal = {\araa},
         year = 1998,
        month = jan,
       volume = {36},
        pages = {189-232},
          doi = {10.1146/annurev.astro.36.1.189},
archivePrefix = {arXiv},
       eprint = {astro-ph/9807187},
 primaryClass = {astro-ph},
       adsurl = {https://ui.adsabs.harvard.edu/abs/1998ARA&A..36..189K}
}

@ARTICLE{Bicknell1994,
       author = {{Bicknell}, Geoffrey V.},
        title = "{On the Relationship between BL Lacertae Objects and Fanaroff-Riley I Radio Galaxies}",
      journal = {\apj},
         year = 1994,
        month = feb,
       volume = {422},
        pages = {542},
          doi = {10.1086/173748},
archivePrefix = {arXiv},
       eprint = {astro-ph/9308033},
 primaryClass = {astro-ph},
       adsurl = {https://ui.adsabs.harvard.edu/abs/1994ApJ...422..542B}
}

@ARTICLE{Canvin2005,
       author = {{Canvin}, J.~R. and {Laing}, R.~A. and {Bridle}, A.~H. and {Cotton}, W.~D.},
        title = "{A relativistic model of the radio jets in NGC 315}",
      journal = {\mnras},
         year = 2005,
        month = nov,
       volume = {363},
       number = {4},
        pages = {1223-1240},
          doi = {10.1111/j.1365-2966.2005.09537.x},
archivePrefix = {arXiv},
       eprint = {astro-ph/0508440},
 primaryClass = {astro-ph},
       adsurl = {https://ui.adsabs.harvard.edu/abs/2005MNRAS.363.1223C}
}

@ARTICLE{Boccardi2021,
       author = {{Boccardi}, B. and {Perucho}, M. and {Casadio}, C. and {Grandi}, P. and {Macconi}, D. and {Torresi}, E. and {Pellegrini}, S. and {Krichbaum}, T.~P. and {Kadler}, M. and {Giovannini}, G. and {Karamanavis}, V. and {Ricci}, L. and {Madika}, E. and {Bach}, U. and {Ros}, E. and {Giroletti}, M. and {Zensus}, J.~A.},
        title = "{Jet collimation in NGC 315 and other nearby AGN}",
      journal = {\aap},
         year = 2021,
        month = mar,
       volume = {647},
          eid = {A67},
        pages = {A67},
          doi = {10.1051/0004-6361/202039612},
archivePrefix = {arXiv},
       eprint = {2012.14831},
 primaryClass = {astro-ph.HE},
       adsurl = {https://ui.adsabs.harvard.edu/abs/2021A&A...647A..67B}
}

@ARTICLE{Park2021,
       author = {{Park}, Jongho and {Hada}, Kazuhiro and {Nakamura}, Masanori and {Asada}, Keiichi and {Zhao}, Guangyao and {Kino}, Motoki},
        title = "{Jet Collimation and Acceleration in the Giant Radio Galaxy NGC 315}",
      journal = {\apj},
         year = 2021,
        month = mar,
       volume = {909},
       number = {1},
          eid = {76},
        pages = {76},
          doi = {10.3847/1538-4357/abd6ee},
archivePrefix = {arXiv},
       eprint = {2012.14154},
 primaryClass = {astro-ph.HE},
       adsurl = {https://ui.adsabs.harvard.edu/abs/2021ApJ...909...76P}
}

@ARTICLE{Kaviraj2014,
       author = {{Kaviraj}, Sugata},
        title = "{The significant contribution of minor mergers to the cosmic star formation budget}",
      journal = {\mnras},
         year = 2014,
        month = jan,
       volume = {437},
       number = {1},
        pages = {L41-L45},
          doi = {10.1093/mnrasl/slt136},
archivePrefix = {arXiv},
       eprint = {1310.0007},
 primaryClass = {astro-ph.CO},
       adsurl = {https://ui.adsabs.harvard.edu/abs/2014MNRAS.437L..41K}
}

@ARTICLE{Liang2006,
       author = {{Laing}, R.~A. and {Canvin}, J.~R. and {Cotton}, W.~D. and {Bridle}, A.~H.},
        title = "{Multifrequency observations of the jets in the radio galaxy NGC315}",
      journal = {\mnras},
         year = 2006,
        month = may,
       volume = {368},
       number = {1},
        pages = {48-64},
          doi = {10.1111/j.1365-2966.2006.10099.x},
archivePrefix = {arXiv},
       eprint = {astro-ph/0601660},
 primaryClass = {astro-ph},
       adsurl = {https://ui.adsabs.harvard.edu/abs/2006MNRAS.368...48L}
}

@ARTICLE{Mandal2024,
       author = {{Mandal}, Ankush and {Mukherjee}, Dipanjan and {Federrath}, Christoph and {Bicknell}, Geoffrey V. and {Nesvadba}, Nicole P.~H. and {Mignone}, Andrea},
        title = "{Probing the role of self-gravity in clouds impacted by AGN-driven winds}",
      journal = {\mnras},
         year = 2024,
        month = jun,
       volume = {531},
       number = {1},
        pages = {2079-2110},
          doi = {10.1093/mnras/stae1295},
archivePrefix = {arXiv},
       eprint = {2405.10005},
 primaryClass = {astro-ph.GA},
       adsurl = {https://ui.adsabs.harvard.edu/abs/2024MNRAS.531.2079M}
}

@ARTICLE{Duggal2024,
       author = {{Duggal}, C. and {O'Dea}, C.~P. and {Baum}, S.~A. and {Labiano}, A. and {Tadhunter}, C. and {Worrall}, D.~M. and {Morganti}, R. and {Tremblay}, G.~R. and {Dicken}, D.},
        title = "{Optical- and UV-continuum Morphologies of Compact Radio Source Hosts}",
      journal = {\apj},
         year = 2024,
        month = apr,
       volume = {965},
       number = {1},
          eid = {17},
        pages = {17},
          doi = {10.3847/1538-4357/ad2513},
archivePrefix = {arXiv},
       eprint = {2309.00110},
 primaryClass = {astro-ph.GA},
       adsurl = {https://ui.adsabs.harvard.edu/abs/2024ApJ...965...17D}
}

@ARTICLE{Bicknell1986,
       author = {{Bicknell}, G.~V.},
        title = "{A Model for the Surface Brightness of a Turbulent Low Mach Number Jet. III. Adiabatic Jets of Arbitrary Density Ratio: Application to NGC 315}",
      journal = {\apj},
         year = 1986,
        month = jun,
       volume = {305},
        pages = {109},
          doi = {10.1086/164232},
       adsurl = {https://ui.adsabs.harvard.edu/abs/1986ApJ...305..109B}
}

%\begin{thebibliography}{99}
%\end{thebibliography}

%%%%%%%%%%%%%%%%%%%%%%%%%%%%%%%%%%%%%%%%%%%%%%%%%%

%%%%%%%%%%%%%%%%% APPENDICES %%%%%%%%%%%%%%%%%%%%%

\appendix

\section{{\sl Swift}-UVOT lightcurve} \label{sec:var_test}
\begin{figure}
\centering
	\includegraphics[width=8cm]{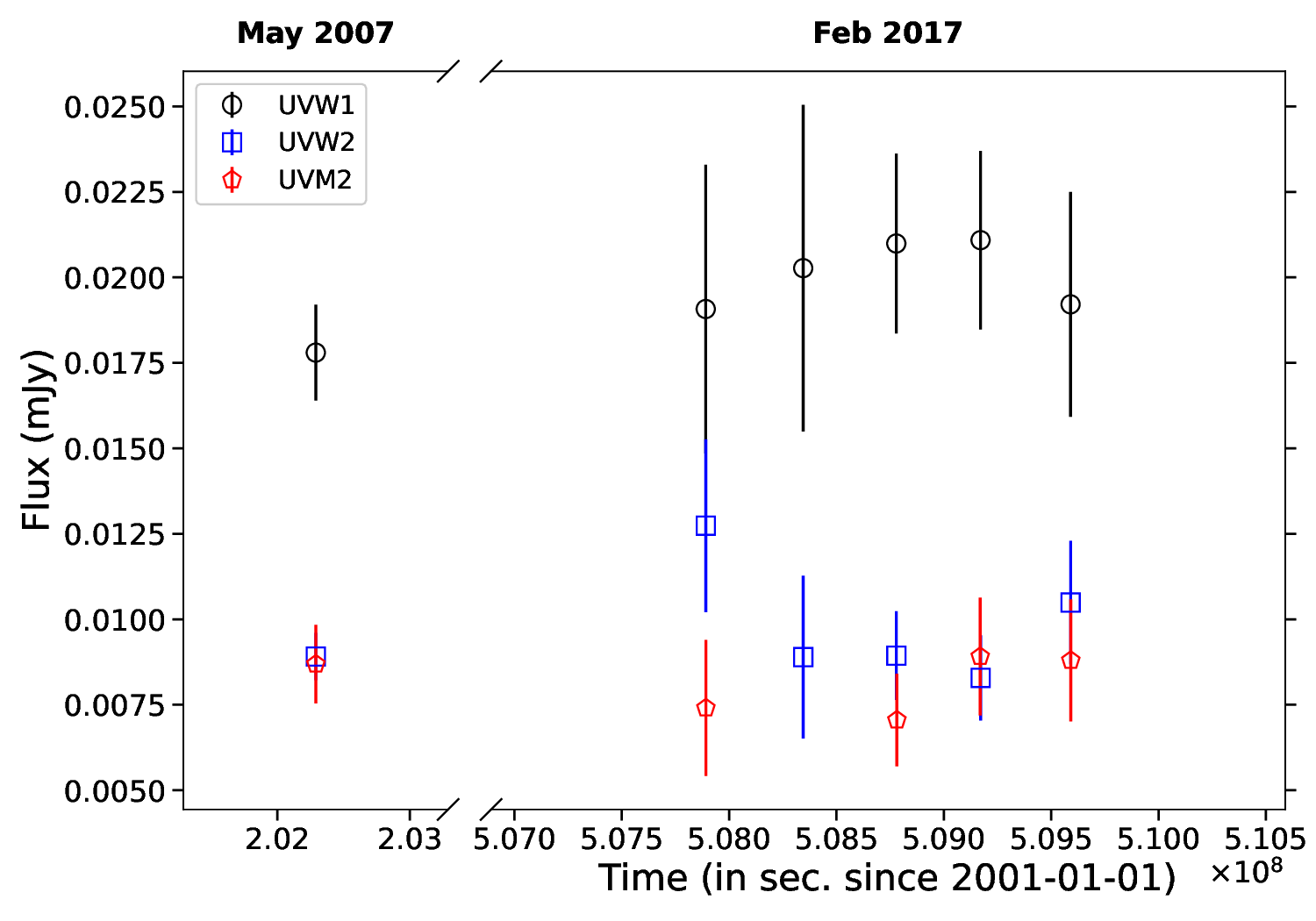}
    \caption{Lightcurve for the UV knot from UVOT observation in {\it uvw1}, {\it uvw2} and {\it uvm2} filters.}
    \label{fig:lc_uvot}
\end{figure}

Lightcurves at the position of the UV knot in {\sl Swift}-UVOT filters are provided in Fig.~\ref{fig:lc_uvot}. Table~\ref{tab:var_test} provides the details of the variability tests \citep[e.g.,][]{Bhattacharya2019}, which show no clear indication of the time variability.

\begin{table}
	\centering
	\caption{Values of different variability indices \label{tab:var_test}}
	
	\begin{tabular}{lcccc} % four columns, alignment for each
		\hline
		Filter & $\chi^2_{\text{red}}$ & RoMS & $\sigma^2_{NXS}$ &$\nu$ \\       
             (1)& (2) & (3) & (4) & (5) \\
		\hline
            {\it uvw1} & $0.40$& $0.56$&$-0.024$ &$-0.019$ \\
            {\it uvw2} & $0.64$ & $0.59$ &$-0.005$ &$0.035$  \\
            {\it uvm2} & $0.34$ & $0.51$ &$-0.03$ &$-0.05$ \\
            \hline
	\end{tabular}

    {\raggedright  \textsc{Notes.--} Col. (1): {\sl Swift}-UVOT filter. Col. (2): Reduced $\chi^2$ \citep{Diego2010}. Col. (3): Robust median statistic \citep{Rose2007}. Col. (4): Normalised excess variance \citep{Nandra1997}. Col. (5): Peak-to-peak variability \citep{Brown1989}. \par}
\end{table}
%%%%%%%%%%%%%%%%%%%%%%%%%%%%%%%%%%%%%%%%%%%%%%%%%%

% Don't change these lines
\bsp	% typesetting comment
\label{lastpage}
\end{document}